\documentclass[final,twoside,onecolumn,letter,12pt]{IEEEtran}
\usepackage{graphicx,cite,amssymb,amsmath,psfrag,subfigure}
\usepackage{graphicx}
\usepackage{amssymb}
\usepackage{amsmath}
\usepackage{cite}
\usepackage{subfigure}
\usepackage{mathrsfs}
\usepackage[displaymath,mathlines]{lineno}
\usepackage{color}
\usepackage{tabulary}
\usepackage{multirow}

\newtheorem{proposition}{Proposition}

\newtheorem{remark}{Remark}

\DeclareMathOperator{\E}{\mathbb{E}}

\newcommand{\EX}[1]{\E\left\{{#1}\right\}}

\newcommand{\PDF}[2]{p_{{#1}}\left({#2}\right)}
\newcommand{\CG}[2]{\mathcal{CN}\left({#1},{#2}\right)}

\newcommand{\B}[1]{{\pmb{#1}}}

\newcommand{\Pu}{p_{\mathrm{u}}}
\newcommand{\Pp}{p_{\mathrm{p}}}
\newcommand{\Eu}{E_{\mathrm{u}}}
\newcommand{\Gu}{p_{\mathrm{u}}}

\def\BibTeX{{\rm B\kern-.05em{\sc i\kern-.025em b}\kern-.08em
    T\kern-.1667em\lower.7ex\hbox{E}\kern-.125emX}}

\begin{document}
%\vspace{-3 cm}
%\linenumbers
\title{
    \hspace{4cm}\\[-0.7cm]
    Energy  and Spectral Efficiency of
    Very Large Multiuser MIMO Systems}
\author{
        Hien Quoc Ngo,
        Erik G. Larsson,
        and
        Thomas L. Marzetta
\thanks{
        H.~Q.\ Ngo and E.~G. Larsson are with the Department of Electrical
    Engineering (ISY), Link\"{o}ping University, 581 83 Link\"{o}ping,
    Sweden
        (Email: nqhien@isy.liu.se; egl@isy.liu.se).
}
\thanks{
        T.~L.\ Marzetta is with Bell Laboratories,
        Alcatel-Lucent, 600 Moutain Avenue,
    Murray Hill, NJ 07974, USA
        (Email: tom.marzetta@alcatel-lucent.com).
}
 \thanks{This work was supported  in part by the Swedish Research Council
(VR), the Swedish
    Foundation for Strategic Research (SSF), and ELLIIT. E. Larsson is a Royal
    Swedish Academy of Sciences (KVA) Research Fellow supported by a
    grant from the Knut and Alice Wallenberg Foundation. %}
%\thanks{
        Parts of this work were presented at the 2011 Allerton Conference on Communication, Control and Computing \cite{NLM:11:ACCCC}.
} }

\markboth{Submitted to the IEEE Transactions on Communications}
        { }

\maketitle

\renewcommand{\baselinestretch}{1.5} \normalsize

% The above \baselinestretch command does not change automatically.
% A font size changing command should be executed to make the new value
% in effect.  I trick this by using \normalsize, which does not change
% the font size!
%%%%%%%%%%%%%%%%%%%%%%%%%%%%%%%%%%%%%%%%%%%%%%%%%%%%%%%%%%%%%%%%%%%%%
\vspace{-1.0cm}

\begin{abstract}

A multiplicity of autonomous terminals simultaneously transmits
data streams to a compact array of antennas. The array uses
imperfect channel-state information derived from transmitted
pilots to extract the individual data streams. The power radiated
by the terminals can be made inversely proportional to the
square-root of the number of base station  antennas with no
reduction in performance. In contrast if perfect channel-state
information were available the power could be made inversely
proportional to the number of antennas. Lower capacity bounds for
maximum-ratio  combining (MRC), zero-forcing (ZF) and minimum
mean-square error (MMSE) detection are derived. A MRC receiver
normally performs worse than ZF and MMSE. However as power levels
are reduced, the cross-talk introduced by the inferior
maximum-ratio receiver eventually falls below the noise level and
this simple receiver becomes a viable option. The tradeoff between
the energy efficiency (as measured in bits/J) and spectral
efficiency (as measured in bits/channel use/terminal) is
quantified. It is shown that the use of moderately large antenna
arrays can improve the spectral and energy efficiency with orders
of magnitude compared to a single-antenna system.

\end{abstract}

\vspace{-0.5 cm}

\begin{keywords}
 Energy efficiency, spectral efficiency, multiuser MIMO,  very large MIMO systems
\end{keywords}
\vspace{-0.5 cm}
%\clearpage
\section{Introduction}

In multiuser multiple-input multiple-output (MU-MIMO) systems, a
base station (BS) equipped with multiple antennas serves a number
of users.   Such systems have attracted much attention for some
time now \cite{GKHCS:07:SPM}. Conventionally, the communication
between the BS and the users is performed by orthogonalizing the
channel so that the BS communicates with each user in separate
time-frequency resources. This is not optimal from an
information-theoretic point of view, and higher rates can be
achieved  if the BS communicates with several users in the same
time-frequency resource \cite{CJKR:10:IT,JAMV:11:WCOM}. However,
complex techniques to mitigate  inter-user interference must then
be used, such as maximum-likelihood multiuser detection on the
uplink \cite{Verdu:89:Al}, or ``dirty-paper coding'' on the
downlink \cite{VT:03:IT,WSS:06:IT}.

Recently, there has been a great deal of interest in MU-MIMO with
\emph{very large antenna arrays} at the BS. Very large arrays can
substantially reduce intracell interference with simple signal
processing \cite{Mar:10:WCOM}. We refer to such systems as ``very
large MU-MIMO systems'' here, and with very large  we mean arrays
comprising say a hundred, or a few hundreds, of antennas,
simultaneously serving tens of users. The design and analysis of
very large MU-MIMO systems is a fairly new subject that is
attracting substantial interest
\cite{Mar:06:ACSSC,Mar:10:WCOM,RPLLMET:11:SPM,HBD:11:ACCCC}. The
vision is that each individual antenna can have a small physical
size, and  be built from inexpensive hardware. With a very large
antenna array, things that were random before start to look
deterministic. As a consequence, the effect of small-scale fading
can be averaged out. Furthermore,  when the number of BS antennas
grows large, the random channel vectors between the users and the
BS become pairwisely orthogonal \cite{RPLLMET:11:SPM}. In the
limit of an infinite number of antennas, with simple matched
filter processing at the BS, uncorrelated noise and intracell
interference disappear completely  \cite{Mar:10:WCOM}.
   Another important
advantage of large MIMO systems is that they enable us to reduce
the transmitted power. On the uplink, reducing the transmit power
of the terminals will drain their batteries slower.  On the
downlink, much of the electrical power consumed by a BS is spent
by power amplifiers and associated circuits and cooling systems
\cite{ComMag}. Hence reducing the emitted RF power would help in
cutting the electricity consumption of the BS.

This paper analyzes the  potential for power savings on the uplink
of very large MU-MIMO systems. We derive new capacity bounds of
the uplink for finite number of BS antennas. These results are
different from recent results in \cite{HCPR:11:APWC} and
\cite{WCSD:10:SPAWC}. In \cite{HCPR:11:APWC} and
\cite{WCSD:10:SPAWC}, the authors derived a deterministic
equivalent of the SINR assuming that the number of transmit
antennas and the number of users go to infinity but their ratio
remains bounded for the downlink of network MIMO systems using a
sophisticated scheduling scheme and MISO broadcast channels using
zero-forcing (ZF) precoding, respectively. While it is well known
that MIMO technology can offer improved power efficiency, owing to
both array gains and diversity effects \cite{TV:05:Book}, we are
not aware of any work that analyzes power efficiency of MU-MIMO
systems with receiver structures that are realistic for very large
MIMO.\footnote{
    After submitting this work, other papers have also addressed
    the tradeoff between spectral and energy efficiency in MU-MIMO
    systems. An analysis related to the one presented here but for
    the downlink was given in \cite{YM:12:JSAC}. However, the analysis of the downlink is
    quantitatively and qualitatively different both in what
    concerns systems aspects and the corresponding the capacity
    bounds.
} We consider both single-cell and multicell systems, but focus on
the analysis of single-cell MU-MIMO systems since: i) the results
are easily comprehensible; ii) it bounds the performance of a
multicell system; and iii) the single-cell performance can be
actually attained if one uses successively less-aggressive
frequency-reuse (e.g., with reuse factor $3$, or $7$). %Given the huge spectral-efficiency
%improvements (factors of 10 - 20, or more) over conventional
%wireless technology, a simple way to deal with multi-cell pilot
%contamination is to sacrifice some of this throughput.
  %The basic question that we address
%is how much the transmitted power of each user can be reduced when
%increasing then number of antennas.
The paper makes the following
specific contributions:
\begin{itemize}
\item We show that, when the number of BS antennas $M$ grows
  without bound, we can reduce the transmitted power of each user
  proportionally to $1/M$ if the BS has perfect channel
  state information (CSI), and proportionally to $1/\sqrt{M}$ if  CSI is estimated
  from uplink pilots.  This
  holds true even when using simple, linear receivers. We also
  derive closed-form expressions of lower bounds on the
uplink
  achievable rates for finite $M$, for the cases of perfect
  and imperfect CSI, assuming MRC,
  ZF, and minimum mean-squared error (MMSE) receivers, respectively. See Section~\ref{Sec:
    Rate}.

\item We study the tradeoff between spectral efficiency and energy
efficiency. For imperfect CSI, in the low transmit power regime,
we can simultaneously increase the spectral-efficiency and
energy-efficiency. We further show that in large-scale MIMO, very
high spectral efficiency can be obtained even with simple MRC
processing at the same time as the transmit power can be cut back
by orders of magnitude and that this holds true even when taking
into account the losses associated with acquiring  CSI from uplink
pilots. MRC also has the advantage that it can be implemented in a
distributed manner, i.e., each antenna performs multiplication of
the received signals with the conjugate of the channel, without
sending the entire baseband signal to the BS for processing. See
Section~\ref{sec: Energy-Spectral}.
\end{itemize}

%\textit{Notation:}   The superscripts $T$, $\ast$, and $H$ stand
%for the transpose, conjugate, and conjugate-transpose,
%respectively. $\left[\B{A}\right]_{ij}$ or $\B{A}_{ij}$ denotes
%the ($i,j$)th entry of a matrix $\B{A}$. The expectation operator
%and the Euclidean norm are denoted by
%$\mathbb{E}\left\{\cdot\right\}$ and $\| \cdot \|$, respectively.
%Finally, we use $\B{z} \sim \CG{0}{\B{\Sigma}}$ to denote a
%circularly symmetric complex Gaussian vector $\B{z}$ with
%covariance matrix $\B{\Sigma}$ and zero mean.

\vspace{-0.5cm}

\section{System Model and Preliminaries} \label{sec: system}

\subsection{MU-MIMO System Model} \label{subsec:Model}

We consider the uplink of a MU-MIMO system. The system  includes
one BS equipped with an array of $M$ antennas that receive data
from $K$ single-antenna users. The nice thing about single-antenna
users is that they are inexpensive, simple, and power-efficient,
and each user still gets typically high throughput. Furthermore,
the assumption that users have single antennas can be considered
as a special case of users having multiple antennas when we treat
the extra antennas as if they were additional autonomous users.
The users transmit their data in the same time-frequency resource.
The $M \times 1$ received vector at the BS is
\begin{align} \label{eq MU-MIMO 1}
    \B{y}
    =
        \sqrt{\Pu}
            \B{G}
            \B{x}
        +
        \B{n}
\end{align}
where $\B{G}$ represents the $M \times K$ channel matrix between
the BS and the $K$ users, i.e., $g_{mk} \triangleq
\left[\B{G}\right]_{mk}$ is the channel coefficient between the
$m$th antenna of the BS and the $k$th user; $\sqrt{\Pu} \B{x}$ is
the $K \times 1$  vector of symbols simultaneously transmitted by
the $K$ users (the average transmitted power of each user is
$\Pu$); and $\B{n}$ is a vector of additive white, zero-mean
Gaussian noise.   We take the noise variance to be $1$, to
minimize notation, but without loss of generality. With this
convention, $\Pu$ has the interpretation of normalized
``transmit'' SNR and is therefore dimensionless. The model
\eqref{eq MU-MIMO 1} also applies to wideband channels handled by
OFDM over restricted intervals of frequency.

The channel matrix $\B{G}$ models independent fast fading,
geometric attenuation, and log-normal shadow fading. The
coefficient $g_{mk}$ can be written as
\begin{align}\label{eq CM 1}
    g_{mk}
    =
        h_{mk} \sqrt{\beta_{k}},~~~ m=1, 2, ..., M
\end{align}
where $h_{mk}$ is the fast fading coefficient from the $k$th user
to the $m$th antenna of the BS. $\sqrt{\beta_{k}}$ models the
geometric attenuation and shadow fading which is assumed to be
independent over $m$ and to be constant over many coherence time
intervals and known a priori. This assumption is reasonable since
the distances between the users and the BS are much larger than
the distance between the antennas, and the value of ${\beta_{k}}$
changes very slowly with time. Then, we have
\begin{align}\label{eq CM 2}
    \B{G}
    =
        \B{H} \B{D}^{1/2}
\end{align}
where $\B{H}$ is the $M \times K$ matrix of fast fading
coefficients between  the $K$ users  and the BS, i.e.,
$\left[\B{H}\right]_{mk}=h_{mk}$, and $\B{D}$ is a $K \times K$
diagonal matrix, where $\left[\B{D}\right]_{kk}=\beta_{k}$.
Therefore, \eqref{eq MU-MIMO 1} can be written as
\begin{align} \label{eq MU-MIMO 2}
    \B{y}
    =
        \sqrt{\Pu}
            \B{H}
            \B{D}^{1/2}
            \B{x}
        +
        \B{n}.
\end{align}
%

%\vspace{-1cm}
\subsection{Review of Some  Results on Very Long Random Vectors}
\label{sec: VeryLongVector}

We   review some  limit results for random vectors
\cite{Cra:70:Book} that will be useful later on.
  Let $\B{p}
\triangleq \left[p_1 ~ ... ~ p_n\right]^T$ and $\B{q}
  \triangleq \left[q_1 ~ ... ~ q_n \right]^T$ be mutually independent $n \times 1$ vectors
  whose elements are i.i.d. zero-mean random variables (RVs) with  $\EX{\left|p_i\right|^2}= \sigma_{p}^2$, and
  $\EX{\left|q_i\right|^2}=\sigma_q^2$, $i=1, ..., n$.
Then from the law of large numbers,
\begin{align} \label{eq Asy 1a}
    \frac{1}{n} \B{p}^H \B{p} \mathop  \to \limits^{a.s.} \sigma_p^2, ~
    \text{and} ~ ~
    %\\ \label{eq Asy 1b}
    \frac{1}{n} \B{p}^H \B{q} \mathop  \to \limits^{a.s.} 0, ~ \text{as} ~
    n \rightarrow \infty.
\end{align}
where $\mathop  \to \limits^{a.s.} $ denotes the almost sure
convergence. Also, from the Lindeberg-L\'{e}vy central limit
theorem,
\begin{align} \label{eq Asy 2}
   \frac{1}{\sqrt{n}} \B{p}^H \B{q} \mathop  \to \limits^d \CG{0}{\sigma_p^2 \sigma_q^2}, ~ \text{as} ~
    n \rightarrow \infty
\end{align}
where $\mathop  \to \limits^d $ denotes convergence in
distribution.

%\item Let $X_1$, $X_2$, ... be a sequence of independent RVs, such
%that $X_i$ has zero mean  and variance $\sigma_i^2$. Further
%assume that the following conditions are satisfied: 1) $s_n^2 =
%\sum_{i=1}^n \sigma_i^2 \rightarrow \infty$, as $n \rightarrow
%\infty$; and 2) $\sigma_i/s_n \rightarrow 0$, as $n \rightarrow
%\infty$. Then by applying the Cram\'er's central limit theorem
%\cite{Cra:70:Book}, we have
%%
%\begin{align} \label{eq Asy 3}
%    \frac{\sum_{i=1}^n X_i}
%         {s_n}
%    \rightarrow
%    \CG{0}{1}, ~ \text{as} ~
%    n \rightarrow \infty.
%\end{align}
%%

%\vspace{-0.5cm}
\subsection{Favorable Propagation}

Throughout the rest of the paper, we assume that the fast fading
coefficients, i.e., the elements of $\B{H}$ are i.i.d.\ RVs with
zero mean and unit variance. Then the conditions in \eqref{eq Asy
1a}--\eqref{eq Asy 2} are satisfied with $\B{p}$ and $\B{q}$ being
any two distinct columns  of $\B{G}$. In this case we have
$$ \frac{\B{G}^H\B{G}}{M} =\B{D}^{1/2} \frac{\B{H}^H \B{H}}{M} \B{D}^{1/2} \approx \B{D},\quad M\gg K $$
and we say that we have \emph{favorable propagation}. Clearly, if
all fading coefficients are i.i.d.\ and zero mean, we have
favorable propagation. Recent channel measurements campaigns have
shown that multiuser MIMO systems with large antenna arrays have
characteristics that approximate the favorable-propagation
assumption fairly well \cite{RPLLMET:11:SPM}, and therefore
provide experimental justification for this assumption.

To understand why favorable propagation is desirable, consider an
$M\times K$ uplink (multiple-access) MIMO channel $\B{H}$, where
$M \geq K$, neglecting for now path loss and shadowing factors in
$\B{D}$. This channel can offer  a sum-rate of
\begin{align}
    R =  \sum_{k = 1}^{K} \log_2 \left( 1 +
    {\Pu } \lambda_k^2 \right)
\end{align}
    where $\Pu$ is the  power spent per terminal and $\{\lambda_k\}_{k=1}^K$ are the singular values of $\B{H}$, see \cite{TV:05:Book}.
If the channel matrix is normalized such that  $|H_{ij}|\sim 1$
(where $\sim$ means equality of the order of magnitude), then
$\sum_{k=1}^K \lambda_k^2 = \|\B{H}\|^2 \approx MK$.  Under this
constraint the rate $R$ is bounded as
\begin{align}
     \log_2 \left(1 + MK\Pu   \right)
\leq R \leq  K   \log_2 \left( 1 + {M} \Pu \right)
\end{align}
The lower bound (left inequality) is satisfied with equality if
$\lambda_1^2=MK$ and $  \lambda_2^2=\cdots=\lambda_K^2=0$ and
corresponds to a rank-one (line-of-sight) channel. The upper bound
(right inequality) is achieved if  $\lambda_1^2
=\cdots=\lambda_K^2=M$. This occurs if the columns of $\B{H}$ are
mutually orthogonal and have the same norm, which is the case when
we have favorable propagation.

\vspace{-0.5cm}
\section{Achievable Rate and Asymptotic ($M \to \infty$) Power Efficiency}\label{Sec: Rate}

By using a large antenna array, we can reduce the transmitted
power of the users as $M$ grows large,  while maintaining a given,
desired quality-of-service. In this section, we quantify this
potential for power decrease, and   derive achievable rates of the
uplink. Theoretically, the BS can use the maximum-likelihood
detector to obtain optimal performance. However, the complexity of
this detector grows exponentially with $K$. The interesting
operating regime is when both $M$ and $K$ are large, but $M$ is
still (much) larger than $K$, i.e., $1\ll K \ll M$. It is known
that in this case, linear detectors  (MRC, ZF and MMSE)  perform
fairly well \cite{Mar:10:WCOM} and therefore we will restrict
consideration to those detectors in this paper.
  We treat the cases of perfect CSI (Section~\ref{sec:pcsi}) and estimated CSI (Section~\ref{sec:icsi}) separately.

\vspace{-0.5cm}
\subsection{Perfect Channel State Information}\label{sec:pcsi}

We first consider the case when the BS has perfect CSI, i.e.\ it
knows $\B{G}$.   Let $\B{A}$ be an $M \times K$ linear detector
matrix which depends on the channel $\B{G}$.  By using the linear
detector, the received signal is separated into streams by
multiplying it with $\B{A}^H$ as follows
\begin{align} \label{eq: linearpro}
    \B{r}
    &=
        {\B{A}}^{H} \B{y}.
\end{align}
We consider three conventional linear detectors MRC, ZF, and MMSE,
i.e.,
\begin{align} \label{eq:condition}
    \B{A}
    =
    \left\{%
\begin{array}{l}
  \B{G} \hspace{3.9cm} \text{for MRC}\\
  \B{G}\left(\B{G}^H \B{G}\right)^{-1} \hspace{2cm} \text{for ZF} \\
  \B{G}\left(\B{G}^H \B{G} + \frac{1}{\Gu}\B{I}_K\right)^{-1} \hspace{0.5cm} \text{for MMSE}\\
\end{array}%
\right.
\end{align}

From \eqref{eq MU-MIMO 1} and \eqref{eq: linearpro}, the received
vector after using the linear detector is given by
\begin{align} \label{eq: GRate PCSI 1}
    \B{r}
    &=
        \sqrt{\Pu}
        \B{A}^H
        \B{G} \B{x}
        +
        \B{A}^H \B{n}.
\end{align}
Let $r_k$ and $x_k$ be the $k$th elements of the $K \times 1$
vectors $\B{r}$ and $\B{x}$, respectively. Then,
\begin{align} \label{eq: GRate PCSI 2}
    {r_k}
    &=
        \sqrt{\Pu}
        \B{a}_k^H
        \B{G} \B{x}
        +
        \B{a}_k^H \B{n}
%    \nonumber
%    \\
    =
        \sqrt{\Pu}
        \B{a}_k^H
        \B{g}_k {x}_k
        +
        \sqrt{\Pu}
        \sum_{i=1, i\neq k}^{K}
        \B{a}_k^H
        \B{g}_i {x}_i
        +
        \B{a}_k^H \B{n}
\end{align}
where $\B{a}_k$ and $\B{g}_k$ are the $k$th columns of the
matrices $\B{A}$ and $\B{G}$, respectively. For a fixed channel
realization $\B{G}$, the noise-plus-interference term is a random
variable with zero mean and variance $\Gu \sum_{i=1, i \neq k}^{K}
|\B{a}_k^H \B{g}_i|^2 + \|\B{a}_k\|^2$. By modeling this term   as
additive Gaussian noise independent of $x_k$ we can obtain a lower
bound on the achievable rate. Assuming further that the channel is
ergodic so that each codeword spans over a large (infinite) number
of realizations of the fast-fading factor of $\B{G}$, the ergodic
achievable uplink rate of the $k$th user is
\begin{align} \label{eq: GRate PCSI 3}
    R_{\mathrm{P},k}
    &=
       \E
        \left\{
       \log_2
            \left(
            1 +
            \frac{
                \Gu |\B{a}_k^H \B{g}_k |^2
                }{
                \Gu \sum_{i=1, i \neq k}^{K}
                |\B{a}_k^H \B{g}_i|^2
                + \|\B{a}_k\|^2
                }
            \right)
       \right\}
\end{align}
%
%where $\Gu \triangleq \Pu/\sigma^2$.
To approach  this capacity lower bound, the message has to be
encoded over many realizations of all sources of randomness that
enter the model (noise and channel). In practice, assuming
wideband operation, this can be achieved by coding over the
frequency domain, using, for example coded OFDM.

\begin{proposition}\label{Thm PerCSI}
Assume that the BS has perfect CSI and that the transmit power of
each user is scaled with $M$ according to $\Pu = \frac{\Eu}{M}$,
where $\Eu$ is fixed. Then,\footnote{
    As mentioned after \eqref{eq MU-MIMO 1}, $\Pu$ has the interpretation of normalized
    transmit SNR, and it is dimensionless. Therefore $\Eu$ is
    dimensionless too.
}
\begin{align} \label{Thm PCSI 1}
    R_{\mathrm{P},k}
    & \rightarrow
        \log_2\left(1+\beta_{k} \Eu\right), M \rightarrow \infty.
\end{align}
\begin{proof}
We give the proof for the case of an MRC receiver. With MRC,
$\B{A}=\B{G}$ so  $\B{a}_k = \B{g}_k$. From \eqref{eq: GRate PCSI
3}, the achievable uplink rate of the $k$th user is
\begin{align} \label{eq: GRate PCSI 5}
    R_{\mathrm{P},k}^{\tt{mrc}}
    &=
       \E
        \left\{
       \log_2
            \left(
            1 +
            \frac{
                \Gu \| \B{g}_k \|^4
                }{
                \Gu \sum_{i=1, i \neq k}^{K}
                |\B{g}_k^H \B{g}_i|^2
                + \|\B{g}_k\|^2
                }
            \right)
       \right\}
%     =
%        \E
%         \left\{
%        \log_2
%             \left(
%             1 +
%             \frac{
%                 \Gu \| \B{g}_k \|^2
%                 }{
%                 \Gu \sum_{i=1, i \neq k}^{K}
%                 | \tilde{g}_i|^2
%                 + 1
%                 }
%             \right)
%        \right\}
\end{align}
%
%By diving the numerator and denominator of the quotient in
%\eqref{eq: GRate PCSI 4} by $\|\B{g}_k\|^2$, we obtain
%%
%\begin{align} \label{eq: GRate PCSI 5}
%    R_{\mathrm{P},k}^{\tt{mrc}}
%    &=
%       \E
%        \left\{
%       \log_2
%            \left(
%            1 +
%            \frac{
%                \Gu \| \B{g}_k \|^2
%                }{
%                \Gu \sum_{i=1, i \neq k}^{K}
%                | \tilde{g}_i|^2
%                + 1
%                }
%            \right)
%       \right\}
%\end{align}
%
Substituting $\Pu = \frac{\Eu}{M}$ into \eqref{eq: GRate PCSI 5},
and using \eqref{eq Asy 1a}, we obtain \eqref{Thm PCSI 1}. By
using the  law of large numbers, we can arrive at the same result
for the ZF and MMSE receivers. Note from  \eqref{eq CM 2} and
\eqref{eq Asy 1a} that when $M$ grows large, $\frac{1}{M}
\B{G}^{H} \B{G}$ tends to $\B{D}$, and hence the ZF and MMSE
filters tend to that of the MRC.
\end{proof}
\end{proposition}

Proposition~\ref{Thm PerCSI} shows that with perfect CSI at the BS
and a large $M$, the performance of a MU-MIMO system with $M$
antennas at the BS and a transmit power per user of $\Eu/M$ is
equal to the performance of a SISO system with transmit power
$\Eu$, without any intra-cell interference and without any fast
fading. In other words, by using a large number of BS antennas, we
can scale down the transmit power proportionally to $1/M$. At the
same time we increase the spectral efficiency $K$ times by
simultaneously serving $K$ users in the same time-frequency
resource.

\subsubsection{Maximum-Ratio Combining} \label{sec: MRC PCSI}

For MRC, from \eqref{eq: GRate PCSI 5}, by the convexity of
$\log_2\left( 1 + \frac{1}{x}\right)$ and using Jensen's
inequality, we obtain the following lower bound on the achievable
rate:
\begin{align} \label{eq: LBRate PCSI 1a}
    R_{\mathrm{P},k}^{\tt{mrc}}
    \geq
    \tilde{R}_{\mathrm{P},k}^{\tt{mrc}}
    &\triangleq
       \log_2
            \left(
            1 +
            \left(
            \E
            \left\{
            \frac{
                \Gu \sum_{i=1, i \neq k}^{K}
                |\B{g}_k^H \B{g}_i|^2
                + \|\B{g}_k\|^2
                }{
            \Gu \| \B{g}_k \|^4
                }
            \right\}
            \right)^{-1}
            \right)
\end{align}

\begin{proposition}\label{Prop PCSI MCR}
With perfect CSI, Rayleigh fading, and $M\geq 2$, the  uplink
achievable rate from the $k$th user for MRC can be lower bounded
as follows:
\begin{align} \label{eq: LBRate PCSI 1}
    \tilde{R}_{\mathrm{P},k}^{\tt{mrc}}
    &=
    \log_2
    \left(
        1
        +
        \frac{
            \Gu \left(M-1 \right) \beta_k
            }{
            \Gu
            \sum_{i=1, i \neq k}^{K}
            \beta_i
            +
            1
            }
    \right)
\end{align}
\begin{proof}
See Appendix~\ref{app:1}.
\end{proof}
\end{proposition}

If $\Pu=\Eu/M$, and $M$ grows without bound, then from \eqref{eq:
LBRate PCSI 1}, we have
\begin{align} \label{Rate PCSI 7}
       \tilde{R}_{\mathrm{P},k}^{\tt{mrc}}
    &=
    \log_2
    \left(
        1
        +
        \frac{
            \frac{\Eu}{M } \left(M-1 \right) \beta_k
            }{
            \frac{\Eu}{M }
            \sum_{i=1, i \neq k}^{K}
            \beta_i
            +
            1
            }
    \right)
%    \nonumber
%    \\
    \rightarrow
       \log_2
            \left(
            1 +
        \beta_k \Eu
            \right), ~ M \rightarrow \infty
\end{align}
Equation \eqref{Rate PCSI 7} shows that the lower bound in
(\ref{eq: LBRate PCSI 1}) becomes equal to the exact limit in
Proposition~\ref{Thm PerCSI} as $M\to\infty$. %This is so because
%when $M\rightarrow \infty$, things that were random before become
%deterministic and hence,  Jensen's inequality will hold with
%equality.

\subsubsection{Zero-Forcing Receiver} \label{sec: ZF PCSI}

With ZF, $\B{A}^H = \left(\B{G}^H\B{G} \right)^{-1}\B{G}^H$, or
$\B{A}^H \B{G} = \B{I}_K$. Therefore,
$
    \B{a}_k^H \B{g}_i
    =
        \delta_{ki},
$
where $\delta_{ki}=1$ when $k=i$ and $0$ otherwise. From
\eqref{eq: GRate PCSI 3}, the uplink rate for the $k$th user is
\begin{align} \label{ZFRate PCSI 2}
    R_{\mathrm{P},k}^{\tt{zf}}
    &=
       \E
        \left\{
       \log_2
            \left(
            1 +
            \frac{
                \Gu
                }{
                \left[\left(\B{G}^H \B{G}\right)^{-1}\right]_{kk}
                }
            \right)
       \right\}.
\end{align}
By using Jensen's inequality, we obtain the following lower bound
on the achievable rate:
\begin{align} \label{ZFRate PCSI 2c}
    R_{\mathrm{P},k}^{\tt{zf}}
    \geq
    \tilde{R}_{\mathrm{P},k}^{\tt{zf}}
    &=
       \log_2
            \left(
            1 +
            \frac{
                \Gu
                }{
                \E  \left\{
                \left[\left(\B{G}^H \B{G}\right)^{-1}\right]_{kk}
                \right\}
                }
            \right)
\end{align}

\begin{proposition}\label{Prop PCSI ZF}
When using ZF, in Rayleigh fading, and provided that $M \geq K+1$,
the achievable uplink rate for the $k$th user   is lower bounded
by
\begin{align} \label{ZFRate PCSI 2b}
    \tilde{R}_{\mathrm{P},k}^{\tt{zf}}
    &=
       \log_2
            \left(
            1 +
            \Gu \left(M-K\right) \beta_k
            \right)
\end{align}
\begin{proof}
See Appendix~\ref{app:2}.
\end{proof}
\end{proposition}

If $\Pu = \Eu/M$, and $M$ grows large, we have
\begin{align} \label{ZFRate PCSI 4} %\setcounter{equation}{33}
    \tilde{R}_{\mathrm{P},k}^{\tt{zf}}
    &=
       \log_2
            \left(
            1 +
            \frac{\Eu}{M } \left(M-K\right) \beta_k
            \right)
%    \nonumber
%    \\ \label{ZFRate PCSI 4b}
    \rightarrow
       \log_2
            \left(
            1 +
       \beta_k \Eu
            \right), ~ M \rightarrow \infty
\end{align}
We can see again from \eqref{ZFRate PCSI 4} that the lower bound
becomes exact for large $M$.

\subsubsection{Minimum Mean-Squared Error Receiver}
For MMSE, the detector matrix $\B{A}$ is
\begin{align} \label{MMSE PCSI 1}
    \B{A}^H
    &=
        \left(
            \B{G}^H\B{G} + \frac{1}{\Gu} \B{I}_{K}
        \right)^{-1}
        \B{G}^H
%    \nonumber
%    \\
    =
        \B{G}^H
        \left(
            \B{G}\B{G}^H + \frac{1}{\Gu} \B{I}_{M}
        \right)^{-1}
\end{align}
Therefore, the $k$th column of $\B{A}$ is given by
\cite{KP:08:WCOM}
\begin{align} \label{MMSE PCSI 2}
    \B{a}_k
    &=
        \left(
            \B{G}\B{G}^H + \frac{1}{\Gu} \B{I}_{M}
        \right)^{-1}
        \B{g}_k
    =
        \frac{
            \B{\Lambda}_k^{-1} \B{g}_k
            }{
            \B{g}_k^H \B{\Lambda}_k^{-1} \B{g}_k
            +1
            }
\end{align}
where $\B{\Lambda}_k \triangleq \sum_{i=1, i \neq k}^{K} \B{g}_i
\B{g}_i^H +  \frac{1}{\Gu} \B{I}_{M}$. Substituting \eqref{MMSE
PCSI 2} into \eqref{eq: GRate PCSI 3}, we obtain the uplink  rate
for user $k$:
\begin{align} \label{MMSERate PCSI 1}
    &R_{\mathrm{P},k}^{\tt{mmse}}
    =
       \E
        \left\{
       \log_2
            \left(
            1 +
            \B{g}_k^H
            \B{\Lambda}_k^{-1}
            \B{g}_k
            \right)
       \right\}
    \mathop = \limits^{(a)}
       \E
        \left\{
       \log_2
            \left(
            \frac{
                1
                }{
                1-\B{g}_k^H\left(\frac{1}{\Pu}\B{I}_M +  \B{G}
                \B{G}^H\right)^{-1} \B{g}_k
                }
            \right)
       \right\}
    \nonumber
    \\
    &=
       \E
        \left\{
       \log_2
            \left(
            \frac{
                1
                }{
                1-\left[\B{G}^H\left(\frac{1}{\Pu}\B{I}_M +  \B{G}
                \B{G}^H\right)^{-1} \B{G}\right]_{kk}
                }
            \right)
       \right\}
    \mathop = \limits^{(b)}
       \E
        \left\{
       \log_2
            \left(
            \frac{
                1
                }{
                \left[\left(\B{I}_K + \Gu \B{G}^H \B{G}\right)^{-1}\right]_{kk}
                }
            \right)
       \right\}
\end{align}
where $(a)$ is obtained directly from \eqref{MMSE PCSI 2}, and
$(b)$ is obtained by using the identity
$$\B{G}^H\left(\frac{1}{\Pu}\B{I}_M + \B{G}
                \B{G}^H\right)^{-1} \B{G} = \left(\frac{1}{\Pu}\B{I}_K +
                \B{G}^H
                \B{G}\right)^{-1} \B{G}^H\B{G} = \B{I}_K- \left(\B{I}_K +
                \Pu \B{G}^H \B{G}\right)^{-1}.$$

By using Jensen's inequality, we obtain the following lower bound
on the
 achievable uplink rate:
\begin{align} \label{BoundMMSE PCSI 1}
    R_{\mathrm{P},k}^{\tt{mmse}} \geq \tilde{R}_{\mathrm{P},k}^{\tt{mmse}}
    &=
       \log_2
            \left(
            1
            +
            \frac{
                1
                }{
                \E\left\{ \frac{1}{\gamma_k} \right\}
                }
            \right)
\end{align}
where
%
%\begin{align} \label{BoundMMSE PCSI 2}
   $ \gamma_k
    =
            \frac{
                1
                }{
                \left[\left(\B{I}_K + \Gu \B{G}^H \B{G}\right)^{-1}\right]_{kk}
                }
            - 1.$
%\end{align}
%
For Rayleigh fading, the exact distribution of $\gamma_k$ can be
found in \cite{GSC:98:COM}. This distribution is analytically
intractable. To proceed, we approximate it with a distribution
which has an analytically tractable form. More specifically, the
PDF of $\gamma_k$ can be approximated by a Gamma distribution as
follows \cite{LPNC:06:IT}:
\begin{align} \label{BoundMMSE PCSI 3}
    \PDF{\gamma_k}{\gamma}
    &=
    \frac{
        \gamma^{\alpha_k -1}
        e^{-\gamma/\theta_k}
        }{
        \Gamma \left(\alpha_k \right)
        \theta_k^{\alpha_k}
        }
\end{align}
where
\begin{align} \label{BoundMMSE PCSI 3b}
    \alpha_k
    &=
        \frac{
            \left(M-K + 1 + \left(K-1 \right) \mu \right)^2
            }{
            M-K + 1 + \left(K-1\right)\kappa
            }, ~
%    \nonumber
%    \\
    \theta_k
    =
        \frac{
            M-K + 1 + \left(K-1\right)\kappa
            }{
            M-K + 1 + \left(K-1 \right) \mu
            }
        \Gu \beta_k
\end{align}
and where $\mu$ and $\kappa$ are determined by  solving following
equations:
\begin{align} \label{BoundMMSE PCSI 3c}
    &\mu
    =
        \frac{
            1
            }{
            K-1
            }
        \sum_{i=1, i \neq k}^{K}
            \frac{
                1
                }{
                M \Gu \beta_i \left(1 - \frac{K-1}{M} + \frac{K-1}{M} \mu\right)
                + 1
                }
    \nonumber
    \\
    &\kappa
    \left(
        1
        +
        \sum_{i=1, i\neq k}^{K}
            \frac{
                \Gu\beta_i
                }{
                \left(M \Gu \beta_i \left(1 - \frac{K-1}{M} + \frac{K-1}{M} \mu\right)
                + 1
                \right)^2
                }
    \right)
    =
        \sum_{i=1, i\neq k}^{K}
            \frac{
                \Gu\beta_i \mu + 1
                }{
                \left(M \Gu \beta_i \left(1 - \frac{K-1}{M} + \frac{K-1}{M} \mu\right)
                + 1
                \right)^2
                }
\end{align}

Using the approximate PDF of $\gamma_k$ given by \eqref{BoundMMSE
PCSI 3}, we have the following proposition.

\begin{proposition}\label{sec: Prop PCSI MMSE}
With perfect CSI, Rayleigh fading, and MMSE, the lower bound on
the   achievable rate for the $k$th user  can be approximated as
\begin{align} \label{BoundMMSE PCSI 4}
    \tilde{R}_{\mathrm{P},k}^{\tt{mmse}}
    &=
       \log_2
            \left(
            1
            +
            \left(\alpha_k -1 \right)
            \theta_k
            \right).
\end{align}
\begin{proof}
Substituting \eqref{BoundMMSE PCSI 3} into \eqref{BoundMMSE PCSI
1}, and using the identity \cite[eq.~(3.326.2)]{GR:07:Book}, we
obtain
\begin{align} \label{BoundMMSE PCSI 4b}
    \tilde{R}_{\mathrm{P},k}^{\tt{mmse}}
    &=
       \log_2
            \left(
            1
            +
            \frac{\Gamma \left(\alpha_k \right)}{\Gamma\left(\alpha_k -1 \right)}
            \theta_k
            \right)
\end{align}
where $\Gamma \left(\cdot\right)$ is the Gamma function. Then,
using $\Gamma\left(x+1 \right) = x \Gamma\left(x\right)$, we
obtain the desired result \eqref{BoundMMSE PCSI 4}.
\end{proof}
\end{proposition}

\begin{remark} \label{re: MMSE}
From \eqref{eq: GRate PCSI 3}, the achievable rate
$R_{\mathrm{P},k}$ can be rewritten as
\begin{align} \label{eq: GRate remark 1}
    R_{\mathrm{P},k}
    &\!=\!
       \E\!
        \left\{\!\!
       \log_2
            \!\left(\!
            1 \!+\!
            \frac{
                |\B{a}_k^H \B{g}_k |^2
                }{
                \B{a}_k^H
                \B{\Lambda}_k
                \B{a}_k
                }
            \!\right)
       \!\!\right\}
%   \nonumber
%   \\
    \!\leq\!
       \E\!
        \left\{\!\!
       \log_2\!
            \left(\!
            1 \!+\!
            \frac{
                \|\B{a}_k^H \B{\Lambda}_k^{1/2} \|^2 \|\B{\Lambda}_k^{-1/2} \B{g}_k \|^2
                }{
                \B{a}_k^H
                \B{\Lambda}_k
                \B{a}_k
                }
            \!\right)
       \!\!\right\}
    =
       \E
        \left\{
       \log_2
            \left(
            1 +
            \B{g}_k^H
            \B{\Lambda}_k^{-1}
            \B{g}_k
            \right)
       \right\}
\end{align}
The inequality is obtained by using Cauchy-Schwarz' inequality,
which holds with  equality  when $\B{a}_k = c
\B{\Lambda}_k^{-1}\B{g}_k$, for any $c \in \mathbb{C}$. This
corresponds to the MMSE detector (see \eqref{MMSE PCSI 2}). This
implies that the MMSE detector is optimal in the sense that it
maximizes the achievable rate given by \eqref{eq: GRate PCSI 3}.
\end{remark}

%\vspace{-1cm}
\subsection{Imperfect Channel State Information}\label{sec:icsi}

In practice, the channel matrix $\B{G}$ has to be estimated at the
BS. The standard way of doing this is to use uplink pilots. A part
of the coherence interval of the channel is then used for the
uplink training. Let $T$ be the length (time-bandwidth product) of
the coherence interval and let $\tau$ be the number of symbols
used for pilots. During the training part of the coherence
interval, all users simultaneously transmit mutually orthogonal
pilot sequences of length $\tau$ symbols.  The pilot sequences
used by the $K$ users can be represented by a $\tau \times K$
matrix $\sqrt{\Pp} \B{\Phi}$ ($\tau \geq K$), which satisfies
$\B{\Phi}^{H}\B{\Phi}=\B{I}_K$, where ${\Pp}\triangleq \tau \Pu$.
Then, the $M \times \tau$ received pilot matrix at the BS is given
by
\begin{align} \label{IPCSI 1}
    \B{Y}_{\mathrm{p}}
    =
        \sqrt{\Pp}
            \B{G}
            \B{\Phi}^{T}
        +
        \B{N}
\end{align}
where ${\B{N}}$ is an $M \times \tau$ matrix with i.i.d.
$\CG{0}{1}$ elements. The MMSE estimate of $\B{G}$ given $\B{Y}$
is
\begin{align} \label{IPCSI 2}
    \hat{\B{G}}
    =
    \frac{1}{\sqrt{\Pp}}
    \B{Y}_{\mathrm{p}}
    \B{\Phi}^{\ast} \tilde{\B{D}}
    =
        \left(
        \B{G}
        +
        \frac{1}{\sqrt{\Pp}} \B{W}
        \right)
        \tilde{\B{D}}
\end{align}
where ${\B{W}} \triangleq \B{N} \B{\Phi}^{\ast}$, and
$\tilde{\B{D}} \triangleq \left(\frac{1}{\Pp} \B{D}^{-1} + \B{I}_K
\right)^{-1}$. Since $\B{\Phi}^{H}\B{\Phi}=\B{I}_K$, $\B{W}$ has
i.i.d.\ $\CG{0}{1}$ elements. Note that our analysis takes into
account the fact that pilot signals cannot take advantage of the
large number of receive antennas since channel estimation has to
be done on a per-receive antenna basis. All results that we
present take this fact into account. Denote by $\B{\mathcal{E}}
\triangleq \hat{\B{G}} - \B{G}$. Then, from \eqref{IPCSI 2}, the
elements of the $i$th column of $\B{\mathcal{E}}$ are RVs with
zero means and variances $\frac{\beta_i }{\Pp \beta_i + 1}$.
Furthermore, owing to the properties of MMSE estimation,
$\B{\mathcal{E}}$ is independent of $\hat{\B{G}}$. The received
vector at the BS can be rewritten as
\begin{align} \label{GRate IPCSI 1}
    \hat{\B{r}}
    &=
        \hat{\B{A}}^H
        \left(
            \sqrt{\Pu}
            \hat{\B{G}}
            \B{x}
        -
            \sqrt{\Pu}
            \B{\mathcal{E}}
            \B{x}
        +
        \B{n}
        \right).
\end{align}
Therefore, after using the linear detector, the received signal
associated with the $k$th user is
\begin{align} \label{GRate IPCSI 2}
    \hat{{r}}_k
    &=
            \sqrt{\Pu}
            \hat{\B{a}}_k^H
            \hat{\B{G}}
            \B{x}
        -
            \sqrt{\Pu}
            \hat{\B{a}}_k^H
            \B{\mathcal{E}}
            \B{x}
        +
        \hat{\B{a}}_k^H
        \B{n}
%    \nonumber
%    \\
    =
            \sqrt{\Pu}
            \hat{\B{a}}_k^H
            \hat{\B{g}}_k
            {x}_k
        +
            \sqrt{\Pu}
            \sum_{i=1, i \neq k}^{K}
            \hat{\B{a}}_k^H
            \hat{\B{g}}_i
            {x}_i
        -
            \sqrt{\Pu}
            \sum_{i=1}^{K}
            \hat{\B{a}}_k^H
            \B{\varepsilon}_i
            {x}_i
        +
        \hat{\B{a}}_k^H
        \B{n}
\end{align}
where $\hat{\B{a}}_k$, $\hat{\B{g}}_i$, and ${\B{\varepsilon}}_i$
are the $i$th columns of $\hat{\B{A}}$, $\hat{\B{G}}$, and
$\B{\mathcal{E}}$, respectively.

Since $\hat{\B{G}}$ and $\B{\mathcal{E}}$ are independent,
$\hat{\B{A}}$ and $\B{\mathcal{E}}$ are independent too. The BS
treats the channel estimate as the true channel, and the part
including the last three terms of \eqref{GRate IPCSI 2} is
considered as interference and noise. Therefore, an achievable
rate of the uplink transmission from the $k$th user is given by
\begin{align} \label{GRate IPCSI 3}
    R_{\mathrm{IP},k}
    &=
       \E
        \left\{
       \log_2
            \left(
            1 +
            \frac{
                \Gu |\hat{\B{a}}_k^H \hat{\B{g}}_k |^2
                }{
                \Gu \sum_{i=1, i \neq k}^{K}
                |\hat{\B{a}}_k^H \hat{\B{g}}_i|^2
                +
                \Gu
                \|\hat{\B{a}}_k\|^2
                \sum_{i=1}^K
                \frac{\beta_i}{\tau\Gu \beta_i + 1}
                +
                \|\hat{\B{a}}_k\|^2
                }
            \right)
       \right\}
\end{align}

Intuitively, if we cut the transmitted power of each user, both
the data signal and the pilot signal suffer from the reduction in
power. Since these signals are multiplied together at the
receiver, we expect that there will be a ``squaring effect''. As a
consequence, we cannot reduce power proportionally to $1/M$ as in
the case of perfect CSI.  The following proposition shows that it
is possible to reduce the power (only) proportionally to
$1/\sqrt{M}$.

\begin{proposition}\label{Thm IPerCSI1}
Assume that the BS has imperfect CSI, obtained by MMSE estimation
from uplink pilots, and that the transmit power of each user is
$\Pu = \frac{\Eu}{\sqrt{M}}$, where $\Eu$ is fixed. Then,
\begin{align} \label{Thm IPCSI1 1}
    R_{\mathrm{IP},k}
    & \rightarrow
        \log_2\left(1+\tau \beta_{k}^2 \Eu^2\right), M \rightarrow \infty.
\end{align}
\begin{proof}
For MRC, substituting $\hat{\B{a}}_k = \hat{\B{g}}_k$ into
\eqref{GRate IPCSI 3}, we obtain the   achievable uplink rate as
\begin{align} \label{MRC IPCSI 2}
    R_{\mathrm{IP},k}^{\tt{mrc}}
    &=
       \E
        \left\{
       \log_2
            \left(
            1 +
            \frac{
                \Gu \|\hat{\B{g}}_k \|^4
                }{
                \Gu \sum_{i=1, i \neq k}^{K}
                |\hat{\B{g}}_k^H \hat{\B{g}}_i|^2
                +
                \Gu
                \|\hat{\B{g}}_k\|^2
                \sum_{i=1}^K
                \frac{\beta_i}{\tau\Gu \beta_i + 1}
                +
                \|\hat{\B{g}}_k\|^2
                }
            \right)
       \right\}
%     \nonumber
%     \\
%     &=
%        \E
%         \left\{
%        \log_2
%             \left(
%             1 +
%             \frac{
%                 \Gu \|\hat{\B{g}}_k \|^2
%                 }{
%                 \Gu \sum_{i=1, i \neq k}^{K}
%                 |\hat{\tilde{g}}_i|^2
%                 +
%                 \Gu
%                 \sum_{i=1}^K
%                 \frac{\beta_i}{\tau\Gu \beta_i + 1}
%                 +
%                 1
%                 }
%             \right)
%        \right\}
\end{align}
%
%Similarly, by diving the numerator and denominator of the quotient
%inside the logarithm function of \eqref{MRC IPCSI 1} by
%$\|\hat{\B{g}}_k\|^2$, we obtain
%%
%\begin{align} \label{MRC IPCSI 2}
%    R_{\mathrm{IP},k}^{\tt{mrc}}
%    &=
%       \E
%        \left\{
%       \log_2
%            \left(
%            1 +
%            \frac{
%                \Gu \|\hat{\B{g}}_k \|^2
%                }{
%                \Gu \sum_{i=1, i \neq k}^{K}
%                |\hat{\tilde{g}}_i|^2
%                +
%                \Gu
%                \sum_{i=1}^K
%                \frac{\beta_i}{\tau\Gu \beta_i + 1}
%                +
%                1
%                }
%            \right)
%       \right\}
%\end{align}
%
% where $\hat{\tilde{g}}_i \triangleq \frac{ \hat{\B{g}}_k^H
% \hat{\B{g}}_i}{\|\hat{\B{g}}_k\|} \sim \CG{0}{\frac{\Pp
% \beta_i^2}{\Pp\beta_i + 1}}$ which is independent of
% $\hat{\B{g}}_k$.
Substituting $\Pu = \Eu/{\sqrt{M}}$ into \eqref{MRC IPCSI 2}, and
again using \eqref{eq Asy 1a} along with the fact that each
element of $\hat{\B{g}}_k$ is a RV with zero mean and variance
$\frac{\Pp \beta_k^2 }{\Pp \beta_k + 1}$, we obtain \eqref{Thm
IPCSI1 1}. We can obtain the limit in \eqref{Thm IPCSI1 1} for ZF
and MMSE in a similar way.
\end{proof}
\end{proposition}

Proposition~\ref{Thm IPerCSI1} implies that with imperfect CSI and
a large $M$, the performance of a MU-MIMO system with an
$M$-antenna array at the BS and with the transmit power per user
set to $\Eu/\sqrt{M}$ is equal to the performance of an
interference-free SISO link with transmit power $\tau \beta_k
\Eu^2$, without fast fading.

\begin{remark}
  From the proof of Proposition~\ref{Thm IPerCSI1}, we see that if
   we cut the transmit power proportionally to $1/M^\alpha$, where
  $\alpha > 1/2$, then the SINR of the uplink transmission from the
  $k$th user will go to zero as $M\to\infty$. This means that
$1/\sqrt{M}$ is the fastest rate at which  we can cut the transmit
  power of each user and still maintain a fixed rate.
\end{remark}

\begin{remark}
  In general, each user can use different transmit powers which depend
  on the geometric attenuation and the shadow fading. This can be done
  by assuming that the $k$th user knows $\beta_k$ and performs power control.
  In this case,  the reasoning leading to Proposition~\ref{Thm
    IPerCSI1} can be extended to show that  to achieve the same rate
  as in a SISO system using transmit power $\Eu$, we must choose the transmit power of the $k$th user to be $\sqrt{\frac{\Eu }{{ M \tau \beta_k}}}$.
\end{remark}

\begin{remark}
It can be seen directly from \eqref{eq: GRate PCSI 5} and
\eqref{MRC IPCSI 2} that the power-scaling laws still hold even
for the most unfavorable propagation case (where $\B{H}$ has rank
one). However, for this case, the multiplexing gains do not
materialize since the intracell interference cannot be cancelled
when $M$ grows without bound.
\end{remark}

\subsubsection{Maximum-Ratio Combining}

By following a similar line of reasoning as in the case of perfect
CSI, we can obtain lower bounds on the achievable rate.

\begin{proposition}\label{sec: Prop IPCSI MRC}
With imperfect CSI, Rayleigh fading, MRC processing,  and for $M
\geq 2$, the achievable uplink rate for the $k$th user   is lower
bounded by
\begin{align} \label{MRC IPCSI 2b}
    \tilde{R}_{\mathrm{IP},k}^{\tt{mrc}}
    &=
       \log_2
            \left(
            1 +
            \frac{
                \tau \Gu^2
                \left(M-1 \right)
                \beta_k^2
                }{
                \Gu
                \left(\tau \Gu\beta_k +1 \right)
                \sum_{i=1, i \neq k}^{K} \beta_i
                +
                \left(\tau +1 \right)\Gu\beta_k
                +
                1
                }
            \right)
\end{align}
\end{proposition}

By choosing $\Pu = \Eu/\sqrt{M}$, we obtain
\begin{align} \label{Rate IPCSI 6}
    &\tilde{R}_{\mathrm{IP},k}^{\tt{mrc}}
    \rightarrow
       \log_2
            \left(
            1 +
                \tau \beta_k^2 \Eu^2
            \right), ~ M \rightarrow \infty
\end{align}
Again, when $M \to \infty$, the asymptotic bound on the rate
equals the exact limit obtained from Proposition~\ref{Thm
IPerCSI1}.

\subsubsection{ZF  Receiver}

For the ZF receiver, we have $\hat{\B{a}}_k^H \hat{\B{g}}_i =
\delta_{ki}$. From \eqref{GRate IPCSI 3}, we obtain the
  achievable uplink rate for the $k$th user as
\begin{align} \label{ZFRate IPCSI 2}
    R_{\mathrm{IP},k}^{\tt{zf}}
    =
       \E
        \left\{
       \log_2
            \left(
            1 +
            \frac{
                \Gu
                }{
                \left(
                    \sum_{i=1}^{K}
                    \frac{\Gu\beta_i }{\tau \Gu \beta_i +1}
                    +
                    1
                \right)
                \left[\left(\hat{\B{G}}^H  \hat{\B{G}}\!\right)^{-1}\right]_{kk}
                }
            \right)
       \right\}.
\end{align}

Following the same derivations as in Section~\ref{sec: ZF PCSI}
for the case of perfect CSI, we  obtain the following  lower bound
on the achievable uplink rate.

\begin{proposition}\label{sec: Prop IPCSI ZF}
With ZF processing using imperfect CSI, Rayleigh fading, and for
$M\geq K+1 $, the
 achievable uplink rate for the $k$th user  is bounded as
\begin{align} \label{ZFRate IPCSI 2b}
    \tilde{R}_{\mathrm{IP},k}^{\tt{zf}}
    =
       \log_2
            \left(
            1 +
            \frac{
                \tau \Gu^2 \left(M-K\right)
                \beta_k^2
                }{
                \left(
                \tau \Gu \beta_k
                +1
                \right)
                    \sum_{i=1}^{K}
                    \frac{\Gu\beta_i }{\tau \Gu \beta_i +1}
                +
                \tau \Gu \beta_k
                +1
                }
            \right).
\end{align}
\end{proposition}

Similarly, with $\Pu=\Eu/\sqrt{M}$, when $M \to \infty$, the
 achievable uplink rate and its  lower  bound  tend to the ones
for MRC (see \eqref{Rate IPCSI 6}), i.e.,
\begin{align} \label{ZFRate IPCSI 4}
    &\tilde{R}_{\mathrm{IP},k}^{\tt{zf}}
    \rightarrow
       \log_2
            \left(
            1 +
                \tau \beta_k^2 \Eu^2
            \right), ~ M \rightarrow \infty
\end{align}
which equals the rate value obtained from Proposition~\ref{Thm
IPerCSI1}.

\subsubsection{MMSE Receiver}
With imperfect CSI, the received vector at the BS can be rewritten
as
\begin{align} \label{eq MU-MIMO IP1}
    \B{y}
    =
        \sqrt{\Pu}
            \hat{\B{G}}
            \B{x}
        -
        \sqrt{\Pu}
            \B{\mathcal{E}}
            \B{x}
        +
        \B{n}
\end{align}
Therefore, for the MMSE receiver, the $k$th column of
$\hat{\B{A}}$ is given by
\begin{align} \label{MMSE IPCSI 1}
    \hat{\B{a}}_k
    &=
        \left(
        \hat{\B{G}}\hat{\B{G}}^H + \frac{1}{\Pu}
        {\tt{Cov}}
        \left(-
        \sqrt{\Pu}
            \B{\mathcal{E}}
            \B{x}
        +
        \B{n}
        \right)
        \right)^{-1}
        \hat{\B{g}}_k
%    \nonumber
%    \\
%    &=
%        \left(
%        \hat{\B{G}}\hat{\B{G}}^H
%        +
%        \left(
%            \sum_{i=1}^K
%                \frac{
%                    \beta_i
%                    }{
%                    \tau \Gu \beta_i
%                    + 1
%                    }
%            +
%            \frac{1}{\Gu}
%        \right)
%        \B{I}_M
%        \right)^{-1}
%        \hat{\B{g}}_k
%    \nonumber
%    \\
    =
     \frac{
            \hat{\B{\Lambda}}_k^{-1} \hat{\B{g}}_k
            }{
            \hat{\B{g}}_k^H \hat{\B{\Lambda}}_k^{-1} \hat{\B{g}}_k
            +1
            }
\end{align}
where ${\tt Cov} \left( \B{a}\right)$ denotes the covariance
matrix of a random vector $\B{a}$, and
\begin{align} \label{MMSE IPCSI 1b}
    \hat{\B{\Lambda}}_k
    \triangleq
        \sum_{i=1, i \neq k}^{K}
            \hat{\B{g}}_i
            \hat{\B{g}}_i^H
            +
        \left(
            \sum_{i=1}^K
                \frac{
                    \beta_i
                    }{
                    \tau \Gu \beta_i
                    + 1
                    }
            +
            \frac{1}{\Gu}
        \right)
        \B{I}_M
\end{align}
Similarly to in Remark~\ref{re: MMSE}, by using Cauchy-Schwarz'
inequality, we can show that the MMSE receiver given by
\eqref{MMSE IPCSI 1} is the optimal detector in the sense that it
maximizes the rate given by \eqref{GRate IPCSI 3}.

Substituting \eqref{MMSE IPCSI 1} into \eqref{GRate IPCSI 3}, we
get the  achievable uplink rate for the $k$th user with  MMSE
receivers as
\begin{align} \label{MMSERate PCSI 1}
    R_{\mathrm{P},k}^{\tt{mmse}}
    &=
       \E
        \left\{
       \log_2
            \left(
            1 +
            \hat{\B{g}}_k^H
            \hat{\B{\Lambda}}_k^{-1}
            \hat{\B{g}}_k
            \right)
       \right\}
%    \nonumber
%    \\
    =
       \E
        \left\{\!\!
       \log_2\!
            \left(
            \frac{
                1
                }{
                \left[\!\left(\B{I}_K
                \!+\!
                \left(
                    \sum_{i=1}^K
                        \frac{
                            \beta_i
                            }{
                            \tau \Gu \beta_i
                            + 1
                            }
                    +
                    \frac{1}{\Gu}
                \right)^{-1}
                 \hat{\B{G}}^H \hat{\B{G}}\right)^{-1}\right]_{kk}
                }
            \right)
       \!\!\right\}.
\end{align}

Again, using an  approximate distribution for the SINR, we can
obtain a lower bound on the achievable uplink rate in closed form.

\begin{proposition}\label{sec: Prop IPCSI MMSE}
With imperfect CSI and Rayleigh fading, the    achievable  rate
for the $k$th user with MMSE processing is approximately lower
bounded as follows:
\begin{align} \label{BoundMMSE IPCSI 1}
    \tilde{R}_{\mathrm{IP},k}^{\tt{mmse}}
    &=
       \log_2
            \left(
            1
            +
            \left(\hat{\alpha}_k -1 \right)
            \hat{\theta}_k
            \right)
\end{align}
where
\begin{align} \label{BoundMMSE IPCSI 1b}
    \hat{\alpha}_k
    &=
        \frac{
            \left(M-K + 1 + \left(K-1 \right) \hat{\mu} \right)^2
            }{
            M-K + 1 + \left(K-1\right)\hat{\kappa}
            }, ~
%    \nonumber
%    \\
    \hat{\theta}_k
    =
        \frac{
            M-K + 1 + \left(K-1\right)\hat{\kappa}
            }{
            M-K + 1 + \left(K-1 \right) \hat{\mu}
            }
        \omega \hat{\beta}_k
\end{align}
where $\omega \triangleq                 \left(
                    \sum_{i=1}^K
                        \frac{
                            \beta_i
                            }{
                            \tau \Gu \beta_i
                            + 1
                            }
                    +
                    \frac{1}{\Gu}
                \right)^{-1}$, $\hat{\beta}_k \triangleq \frac{\tau \Gu \beta_k^2}{\tau \Gu \beta_k + 1}$, $\hat{\mu}$ and $\hat{\kappa}$ are obtained by
using following equations:
\begin{align} \label{BoundMMSE IPCSI 1c}
    &\hat{\mu}
    =
        \frac{
            1
            }{
            K-1
            }
        \sum_{i=1, i \neq k}^{K}
            \frac{
                1
                }{
                M \omega \hat{\beta}_i \left(1 - \frac{K-1}{M} + \frac{K-1}{M} \hat{\mu}\right)
                + 1
                }
    \nonumber
    \\
    &\hat{\kappa}
    \left(
        1
        +
        %\frac{1}{K-1}
        \sum_{i=1, i\neq k}^{K}
            \frac{
                \omega \hat{\beta}_i
                }{
                \left(M \omega \hat{\beta}_i \left(1 - \frac{K-1}{M} + \frac{K-1}{M} \hat{\mu}\right)
                + 1
                \right)^2
                }
    \right)
    =
        \sum_{i=1, i\neq k}^{K}
            \frac{
                \omega \hat{\beta}_i \hat{\mu} + 1
                }{
                \left(M \omega \hat{\beta}_i \left(1 - \frac{K-1}{M} + \frac{K-1}{M} \hat{\mu}\right)
                + 1
                \right)^2
                }
\end{align}
\end{proposition}

   Table~\ref{table:1} summarizes the
lower bounds on the achievable rates for linear receivers derived
in this section, distinguishing between the cases  of perfect and
imperfect CSI, respectively.

We have considered a \emph{single-cell} MU-MIMO system.  This
simplifies the analysis, and it gives us important insights into
how power can be scaled with the number of antennas in  very large
MIMO systems. A natural question is to what extent this
power-scaling law still holds for \emph{multicell} MU-MIMO
systems. Intuitively, when we reduce the transmit power of each
user, the effect of interference from other cells also reduces and
hence, the SINR will stay unchanged. Therefore we will have the
same power-scaling law as in the single-cell scenario. The next
section explains this argument in more detail.

\subsection{Power-Scaling Law for Multicell MU-MIMO Systems} \label{sec: App Multicell}

We will use the MRC for our analysis. A similar analysis can be
performed for the
 ZF and MMSE detectors.
Consider the uplink of a multicell MU-MIMO system with $L$ cells
sharing the same frequency band. Each cell includes one BS
equipped with $M$ antennas and $K$ single-antenna users. The $M
\times 1$ received vector at the $l$th BS is given by
\begin{align} \label{eq MMU-MIMO 1}
    \B{y}_l
    =
        \sqrt{p_{\mathrm{u}}}
        \sum_{i=1}^{L}
            \B{G}_{li}
            \B{x}_i
        +
        \B{n}_l
\end{align}
where $\sqrt{p_{\mathrm{u}}} \B{x}_i$ is the $K \times 1$
transmitted vector of $K$ users in the $i$th cell; $\B{n}_l$ is an
AWGN vector, $\B{n}_l \sim \CG{0}{\B{I}_M}$; and $\B{G}_{li}$ is
the $M \times K$ channel matrix between the $l$th BS and the $K$
users in the $i$th cell. The channel matrix $\B{G}_{li}$ can be
represented as
\begin{align} \label{eq MMU-MIMO 2}
    \B{G}_{li}
    =
        \B{H}_{li}
        \B{D}_{li}^{1/2}
\end{align}
where $\B{H}_{li}$ is the fast fading matrix between the $l$th BS
and the $K$ users in the $i$th cell whose elements have zero mean
and unit variance; and $\B{D}_{li}$ is a $K \times K$ diagonal
matrix, where $\left[\B{D}_{li}\right]_{kk} = \beta_{lik}$, with
$\beta_{lik}$ represents the large-scale fading between the $k$th
user in the $i$ cell and the $l$th BS.

\subsubsection{Perfect CSI}

With perfect CSI, the received signal at the $l$th BS after using
MRC is given by
\begin{align} \label{eq MMU-MIMO 3}
    \B{r}_{l}
    &=
        \B{G}_{ll}^{H}
        \B{y}_l
%    \nonumber
%    \\
    =
        \sqrt{\Pu}
        \B{G}_{ll}^{H}
        \B{G}_{ll}
        \B{x}_l
        +
        \sqrt{\Pu}
        \sum_{i=1, i \neq l}^{L}
            \B{G}_{ll}^{H}
            \B{G}_{li}
            \B{x}_i
        + \B{G}_{ll}^{H} \B{n}_l
\end{align}
With $\Pu = \frac{\Eu}{M}$, \eqref{eq MMU-MIMO 3} can be rewritten
as
\begin{align} \label{eq MMU-MIMO 4}
    \frac{1}{\sqrt{M}}\B{r}_{l}
    &=
        \sqrt{\Eu}
        \frac{\B{G}_{ll}^{H}
        \B{G}_{ll}}{M}
        \B{x}_l
        +
        \sqrt{\Pu}
        \sum_{i=1, i \neq l}^{L}
            \frac{\B{G}_{ll}^{H}
            \B{G}_{li}}{M}
            \B{x}_i
        + \frac{1}{\sqrt{M}}
        \B{G}_{ll}^{H} \B{n}_l
\end{align}
From \eqref{eq Asy 1a}--\eqref{eq Asy 2}, when $M$ grows large,
the interference from other cells disappears.  More precisely,
\begin{align} \label{eq MMU-MIMO 5}
    \frac{1}{\sqrt{M}}
    {\B{r}_l}
    \rightarrow
        \sqrt{\Eu}
            \B{D}_{ll}
            \B{x}_l
        +
        \B{D}_{ll}^{1/2}
        \tilde{\B{n}}_l
\end{align}
where $\tilde{\B{n}}_l \sim \CG{0}{\B{I}}$. Therefore, the SINR of
the uplink transmission from the $k$th user in the $l$th cell
converges to a constant value when $M$ grows large, more precisely
\begin{align} \label{eq MMU-MIMO 6}
    \mathsf{SINR}_{l,k}^{\mathrm{P}}
    \to
        \beta_{llk}
        \Eu, ~
    \text{as} ~ M\to \infty
\end{align}
This means that the power scaling law derived for single-cell
systems is valid in multicell systems too.

\subsubsection{Imperfect CSI}

In this case, the channel estimate from the uplink pilots is
contaminated by interference from other cells. The MMSE channel
estimate of the channel matrix $\B{G}_{ll}$ is given by
\cite{HBD:11:ACCCC}
\begin{align} \label{eq MMSE 5}
    \hat{\B{G}}_{ll}
    &=
        \left(
        \sum_{i =1}^{L}
            \B{G}_{li}
        +
        \frac{1}{\sqrt{p_{\mathrm{p}}}}{\B{W}}_{l}
        \right)
        \tilde{\B{D}}_{ll}
\end{align}
where $\tilde{\B{D}}_{ll}$ is a diagonal matrix where the $k$th
diagonal element $\left[\tilde{\B{D}}_{ll}\right]_{kk}
=\beta_{llk}
        \left(
        \sum_{i = 1}^{L}
            \beta_{lik}
        +
        \frac{1}{p_{\mathrm{p}}}
        \right)^{-1}$. The received signal at the $l$th BS after using
MRC is given by
\begin{align} \label{eq MU-MIMO ICSI 1}
    \hat{\B{r}}_{l}
    &=
        \hat{\B{G}}_{ll}^{H}
        \B{y}_l
%    \nonumber
%    \\
    =
        \tilde{\B{D}}_{ll}
        \left(
        \sum_{i =1}^{L}
            \B{G}_{li}
        +
        \frac{1}{\sqrt{p_{\mathrm{p}}}}{\B{W}}_{l}
        \right)^H
        \left(
        \sqrt{p_{\mathrm{u}}}
        \sum_{i=1}^{L}
            \B{G}_{li}
            \B{x}_i
        +
        \B{n}_l
        \right)
\end{align}
With $\Pu = \Eu/\sqrt{M}$, we have
\begin{align} \label{eq MU-MIMO ICSI 2}
    \frac{1}{M^{3/4}}
    \tilde{\B{D}}_{ll}^{-1}
    \hat{\B{r}}_{l}
    &=
        \sqrt{\Eu}
        \sum_{i=1}^{L}
        \sum_{j=1}^{L}
        \frac{\B{G}_{li}^H \B{G}_{lj}}{M}
        \B{x}_j
        +
        \sum_{i=1}^{L}
        \frac{\B{G}_{li}^H \B{n}_l}{M^{3/4}}
        +
        \frac{1}{\sqrt{\tau}}
        \sum_{i=1}^{L}
        \frac{\B{W}_{l}^H \B{G}_{li}}{M^{3/4}}
        \B{x}_i
        +
        \frac{1}{\sqrt{\tau\Eu}}
        \frac{\B{W}_{l}^H \B{n}_{l}}{M^{1/2}}
\end{align}
By using \eqref{eq Asy 1a} and \eqref{eq Asy 2}, as $M$ grows
large, we obtain
\begin{align} \label{eq MU-MIMO ICSI 3}
    \frac{1}{M^{3/4}}
    \tilde{\B{D}}_{ll}^{-1}
    \hat{\B{r}}_{l}
    &\to
        \sqrt{\Eu}
        \sum_{i=1}^{L}
        \B{D}_{li}
        \B{x}_i
        +
        \frac{1}{\sqrt{\tau\Eu}}
        \tilde{\B{w}}_l
\end{align}
where $\tilde{\B{w}}_l \sim \CG{0}{\B{I}_M}$. Therefore, the
asymptotic SINR of the uplink from the $k$th user in the $l$th
cell is
\begin{align} \label{eq MMU-MIMO 10}
    \mathsf{SINR}_{l,k}^{\mathrm{IP}}
    \rightarrow
        \frac{\tau \beta_{llk}^2 \Eu^2}{\tau \sum_{i \neq l}^{L}\beta_{lik}^2 \Eu^2 + 1}, ~ \text{as} ~ M \rightarrow \infty.
\end{align}
We can see that the $1/\sqrt{M}$ power-scaling law still holds.
Furthermore, transmission from users in other cells constitutes
residual interference. The reason is that the pilot reuse gives
pilot-contamination-induced inter-cell interference which grows
with $M$ at the same rate as the desired signal.

%\vspace{-0.5cm}
\section{Energy-Efficiency versus Spectral-Efficiency Tradeoff}
\label{sec: Energy-Spectral}

  The energy-efficiency (in bits/Joule) of a system is
defined as the spectral-efficiency (sum-rate in bits/channel use)
divided by the transmit power expended (in Joules/channel use).
Typically, increasing the spectral efficiency is associated with
increasing the power and hence, with decreasing the
energy-efficiency. Therefore, there is a fundamental tradeoff
between the energy efficiency and the spectral efficiency.
However, in one operating regime it is possible to jointly
increase the energy and spectral efficiencies, and in this regime
there is no tradeoff. This may appear a bit counterintuitive at
first, but it falls out from the analysis in Section~\ref{sec:
SingleCell EE}. Note, however, that this effect occurs in an
operating regime that is probably of less interest in practice.

In this section, we study the energy-spectral efficiency tradeoff
for the uplink of MU-MIMO systems using linear receivers at the
BS. Certain activities (multiplexing to many users rather than
beamforming to a single user and increasing the number of service
antennas) can simultaneously benefit both the spectral-efficiency
and the radiated energy-efficiency. Once the number of service
antennas is set, one can adjust other system parameters (radiated
power, numbers of users, duration of pilot sequences) to obtain
increased spectral-efficiency at the cost of reduced
energy-efficiency, and vice-versa. This should be a desirable
feature for service providers: they can set the operating point
according to the current traffic demand (high energy-efficiency
and low spectral-efficiency, for example, during periods of low
demand).

%\vspace{-0.5cm}
\subsection{Single-Cell MU-MIMO Systems} \label{sec: SingleCell EE}
We define the  spectral efficiency for perfect and imperfect CSI,
respectively, as follows
\begin{align}\label{eq: ES 1}
    &R_{\mathrm{P}}^{A}
    =
        \sum_{k=1}^{K} \tilde{R}_{\mathrm{P},k}^{A}, ~ \text{and}
        ~
%    \\ \label{eq: ES 1b}
    R_{\mathrm{IP}}^{A}
    =
        \frac{T-\tau}{T}
        \sum_{k=1}^{K} \tilde{R}_{\mathrm{IP},k}^{A}
\end{align}
where $A \in \left\{\tt{mrc, zf, mmse} \right\}$  corresponds to
MRC, ZF and MMSE, and $T$ is the coherence interval in symbols.
The energy-efficiency for perfect and imperfect CSI is defined as
\begin{align} \label{eq: ES 2a}
\eta_{\mathrm{P}}^{A} = \frac{1}{\Pu} R_{\mathrm{P}}^{A}, ~
\text{and} ~
%\\  \label{eq: ES 2}
\eta_{\mathrm{IP}}^{A} = \frac{1}{\Pu} R_{\mathrm{IP}}^{A}
\end{align}
For analytical tractability, we ignore the effect of the
large-scale fading here, i.e., we set $\B{D}=\B{I}_K$.  Also, we
only consider MRC and ZF receivers.\footnote{
    When $M$ is large, the performance of the MMSE
    receiver is very close to that of the  ZF receiver (see
    Section~\ref{Sec:Results}). Therefore, the insights on energy versus spectral
    efficiency obtained from studying the performance
    of ZF can be used to draw conclusions about MMSE as well.
}

For perfect CSI,  it is straightforward to show from \eqref{eq:
LBRate PCSI 1}, \eqref{ZFRate PCSI 2b}, and \eqref{eq: ES 2a} that
when the spectral efficiency increases, the energy efficiency
 decreases.  For imperfect CSI, this is not always so, as we shall see next.
 In what follows, we focus on the case of imperfect CSI since this is the case of interest in practice.

\subsubsection{Maximum-Ratio Combining} From \eqref{MRC IPCSI 2b},
the spectral efficiency and energy efficiency with MRC processing
are   given by
\begin{align}\label{eq: ES MRC1}
    &R_{\mathrm{IP}}^{\tt mrc}
    =
        \frac{T-\tau}{T}
        K
        \log_2
            \left(
                1
                +
                \frac{
                    \tau \left(M-1\right) \Pu^2
                    }{
                    \tau\left(K-1 \right)\Pu^2
                    +
                    \left(
                        K+\tau
                    \right)\Pu
                    + 1
                    }
            \right), ~ \text{and} ~
%    \nonumber
%    \\
    \eta_{\mathrm{IP}}^{\tt mrc} = \frac{1}{\Pu} R_{\mathrm{IP}}^{\tt mrc}
\end{align}
We have
\begin{align}\label{eq: ES MRC2}
    \lim_{\Pu \to 0} \eta_{\mathrm{IP}}^{\tt mrc}
        &=
        \lim_{\Pu \to 0}
            \frac{1}{\Pu} R_{\mathrm{IP}}^{\tt mrc}
%        \nonumber
%        \\
        %\mathop  = \limits^{(a)}
        =
        \lim_{\Pu \to 0}
                \frac{T-\tau}{T}
                K
                \frac{
                    \left(\log_2 e \right)
                    \tau \left(M-1\right) \Pu
                    }{
                    \tau\left(K-1 \right)\Pu^2
                    +
                    \left(
                        K+\tau
                    \right)\Pu
                    + 1
                    }
        =0
\end{align}
and
\begin{align}\label{eq: ES MRC3}
    \lim_{\Pu \to \infty} \eta_{\mathrm{IP}}^{\tt mrc}
        &=
        \lim_{\Pu \to \infty}
            \frac{1}{\Pu} R_{\mathrm{IP}}^{\tt mrc}
          =0
\end{align}
%
%where $(a)$ is obtained by using the fact that $\lim_{x \to 0}
%\frac{\ln \left(1+ x\right)}{x} = 1$.

Equations \eqref{eq: ES MRC2} and \eqref{eq: ES MRC3} imply that
for  low $\Pu$, the energy efficiency increases when $\Pu$
increases, and for  high $\Pu$ the energy efficiency decreases
when $\Pu$ increases. Since $\frac{\partial R_{\mathrm{IP}}^{\tt
mrc}}{\partial \Pu}> 0$, $\forall \Pu > 0$, $R_{\mathrm{IP}}^{\tt
mrc}$ is a monotonically increasing function of $\Pu$. Therefore,
at low $\Pu$ (and hence at low spectral efficiency), the energy
efficiency increases as the spectral efficiency increases and vice
versa at high $\Pu$. The reason is that, the spectral efficiency
 suffers from   a ``squaring effect'' when the received data signal is multiplied with the
 received pilots. Hence,
at $\Pu \ll 1$, the spectral-efficiency behaves as $\sim\Pu^2$. As
a consequence, the energy efficiency (which is defined as the
spectral efficiency divided by $\Pu$) increases linearly with
$\Pu$. In more detail, expanding the rate in a Taylor series for
$\Pu \ll 1$, we obtain
\begin{align}\label{eq: ES MRC4b}
    R_{\mathrm{IP}}^{\tt mrc}
    &\approx
    \left.R_{\mathrm{IP}}^{\tt mrc}\right|_{\Pu=0}
    +
    \left.\frac{
        \partial R_{\mathrm{IP}}^{\tt mrc}
        }{
        \partial \Pu
        }\right|_{\Pu=0}
    \Pu
    +
    \frac{1}{2}
     \left.\frac{
        \partial^2 R_{\mathrm{IP}}^{\tt mrc}
        }{
        \partial \Pu^2
        }\right|_{\Pu=0}
    \Pu^2
%    \nonumber
%    \\
    =
        \frac{T-\tau}{T}
        K
        \log_2
            \left(e
            \right)
        \tau \left(M-1\right) \Pu^2
\end{align}
This gives  the following relation between the spectral efficiency
and energy efficiency at $\Pu \ll 1$:
\begin{align}\label{eq: ES MRC4c}
    \eta_{\mathrm{IP}}^{\tt mrc}
    &=
        \sqrt{
        \frac{T-\tau}{T}
        K
        \log_2
            \left(e
            \right)
        \tau \left(M-1\right) R_{\mathrm{IP}}^{\tt mrc}
        }
\end{align}
We can see that when $\Pu \ll 1$, by doubling the spectral
efficiency, or by doubling $M$, we can increase the energy
efficiency by $1.5$ dB.

\subsubsection{Zero-Forcing Receiver} From \eqref{ZFRate IPCSI 2b},
the spectral efficiency and energy efficiency for ZF are given by
\begin{align}\label{eq: ES ZF1}
    &R_{\mathrm{IP}}^{\tt zf}
    =
        \frac{T-\tau}{T}
        K
        \log_2
            \left(
                1
                +
                \frac{
                    \tau \left(M-K\right) \Pu^2
                    }{
                    \left(
                        K+\tau
                    \right)\Pu
                    + 1
                    }
            \right), ~ \text{and} ~
%    \nonumber
%    \\
    \eta_{\mathrm{IP}}^{\tt zf} = \frac{1}{\Pu} R_{\mathrm{IP}}^{\tt zf}
\end{align}
Similarly to in the analysis of MRC, we can show that at low
transmit power $\Pu$, the energy efficiency increases when the
spectral efficiency increases. In the  low-$\Pu$ regime, we obtain
the following   Taylor series expansion
\begin{align}\label{eq: ES ZF1b}
    R_{\mathrm{IP}}^{\tt zf}
    &\approx
        \frac{T-\tau}{T}
        K
        \log_2
            \left(e
            \right)
        \tau \left(M-K\right) \Pu^2, ~ \text{for $\Pu \ll 1$}
\end{align}
Therefore,
\begin{align}\label{eq: ES ZF1c}
    \eta_{\mathrm{IP}}^{\tt zf}
    &=
        \sqrt{
        \frac{T-\tau}{T}
        K
        \log_2
            \left(e
            \right)
        \tau \left(M-K\right) R_{\mathrm{IP}}^{\tt zf}
        }
\end{align}
Again, at $\Pu \ll 1$, by doubling $M$ or $R_{\mathrm{IP}}^{\tt
zf}$, we can increase the energy efficiency by $1.5$ dB.

\subsection{Multicell MU-MIMO Systems}
In this section, we derive expressions for the energy-efficiency
and spectral-efficiency for a multicell system. These are used for
the simulation in the Section~\ref{Sec:Results}. Here, we consider
a simplified channel model, i.e., $\B{D}_{ll} = \B{I}_K$, and
$\B{D}_{li}=\beta \B{I}_K$, where $\beta \in \left[ 0, 1\right]$
is an intercell interference factor. Note that from \eqref{eq MMSE
5}, the estimate of the channel between the $k$th user in the
$l$th cell and the $l$th BS is given by
\begin{align} \label{eq Muticell 1}
    \hat{\B{g}}_{llk}
%    &=
%        \left(
%        \left(L-1\right)\beta + 1 + \frac{1}{\Pp}
%        \right)^{-1}
%        \left(
%        \sum_{i =1}^{L}
%            \B{g}_{lik}
%        +
%        \frac{1}{\sqrt{p_{\mathrm{p}}}}{\B{w}}_{lk}
%        \right)
%    \nonumber
%    \\
    &=
        \left(
        \left(L-1\right)\beta + 1 + \frac{1}{\Pp}
        \right)^{-1}
        \left(
        \B{h}_{llk}
        +
        \sum_{i \neq k}^{L}
            \sqrt{\beta}\B{h}_{lik}
        +
        \frac{1}{\sqrt{p_{\mathrm{p}}}}{\B{w}}_{lk}
        \right)
\end{align}
The term $\sum_{i \neq k}^{L}\sqrt{\beta}\B{h}_{lik}$ represents
the pilot contamination, therefore $\frac{\sum_{i \neq
k}^{L}\E\left\{\|\sqrt{\beta}\B{h}_{lik}\|^2\right\}}{\E\left\{\|\B{h}_{llk}\|^2\right\}}
= \beta\left(L-1\right)$ can be considered as the effect of pilot
contamination.

Following a similar derivation as in the case of single-cell
MU-MIMO systems, we obtain the spectral efficiency and energy
efficiency for imperfect CSI with MRC and ZF receivers,
respectively, as follows:
\begin{align}\label{eq: ES Multicell MRC1}
    &R_{\mathrm{mul}}^{\tt mrc}
    \!=\!
        \frac{T\!-\!\tau}{T}
        K
        \log_2\!\!
            \left(\!\!
                1
                \!+\!
                \frac{
                    \tau \left(M-1\right) \Pu^2
                    }{
                    \tau\left(\!
                        K \bar{L}^2-1
                        \!+\!
                        \beta\left(\!\bar{L}\!-\!1\!\right)
                        \left(\!M\!-\!2\!\right)
                        \!\right)\Pu^2
                    \!+\!
                    \bar{L}\left(\!
                        K\!+\!\tau
                    \!\right)\Pu
                    \!+ \!1
                    }
            \!\right), \text{and} ~
%    \nonumber
%    \\
    \eta_{\mathrm{mul}}^{\tt mrc} = \frac{1}{\Pu} R_{\mathrm{IP}}^{\tt mrc}
%\end{align}
%%
%%
%\begin{align}\label{eq: ES Multicell ZF1}
\\ \label{eq: ES Multicell ZF1}
    &R_{\mathrm{mul}}^{\tt zf}
    =
        \frac{T-\tau}{T}
        K
        \log_2
            \left(
                1
                +
                \frac{
                    \tau \left(M-K\right) \Pu^2
                    }{
                    \tau K\left(\bar{L}^2-\bar{L}\beta + \beta -1 \right)
                    \Pu^2
                    +
                    \bar{L} \left(
                        K+\tau
                    \right)\Pu
                    + 1
                    }
            \right), ~ \text{and} ~
%    \nonumber
%    \\
    \eta_{\mathrm{IP}}^{\tt zf} = \frac{1}{\Pu} R_{\mathrm{ml}}^{\tt zf}
\end{align}
where $\bar{L} \triangleq \left(L-1\right)\beta + 1$. The
principal complexity in the derivation is the correlation between
pilot-contaminated channel estimates.

We can see that the spectral efficiency is a decreasing function
of $\beta$ and $L$. Furthermore, when $L=1$, or $\beta = 0$, the
results \eqref{eq: ES Multicell MRC1} and \eqref{eq: ES Multicell
ZF1} coincide with \eqref{eq: ES MRC1} and \eqref{eq: ES ZF1} for
single-cell MU-MIMO systems.

\section{Numerical Results} \label{Sec:Results}
\subsection{Single-Cell MU-MIMO Systems}
We consider a hexagonal cell with a radius (from center to vertex)
of $1000$ meters. The users are located uniformly at random in the
cell and we assume that no user is closer to the BS than
$r_\mathrm{h} = 100$ meters. The large-scale fading  is modelled
via $\beta_k=z_k/(r_k/r_\mathrm{h})^\nu$, where $z_k$ is a
log-normal random variable with standard deviation
$\sigma_{\mathrm{shadow}}$, $r_k$ is the distance between the
$k$th user and the BS, and $\nu$ is the path loss exponent. For
all examples, we choose $\sigma_{\mathrm{shadow}} = 8$ dB, and
$\nu=3.8$.

We assume that the transmitted data are modulated with OFDM. Here,
we choose parameters that resemble those of LTE standard: an OFDM
symbol duration of $T_{\mathrm{s}} = 71.4 \mu$s, and a useful
symbol duration of $T_{\mathrm{u}} = 66.7 \mu$s. Therefore, the
guard interval length is $T_{\mathrm{g}} = T_{\mathrm{s}} -
T_{\mathrm{u}} = 4.7\mu$s. We choose the channel coherence time to
be $T_{\mathrm{c}}=1$ ms. Then, $T =
\frac{T_{\mathrm{c}}}{T_{\mathrm{s}}}
\frac{T_{\mathrm{u}}}{T_{\mathrm{g}}} = 196$, where
$\frac{T_{\mathrm{c}}}{T_{\mathrm{s}}} = 14$ is the number of OFDM
symbols in a $1$ ms coherence interval, and
$\frac{T_{\mathrm{u}}}{T_{\mathrm{g}}} =14$ corresponds to the
``frequency smoothness interval" \cite{Mar:10:WCOM}. %For all
%examples, we computed the spectral efficiency using \eqref{eq: ES
%1}.

%\vspace{-0.5cm}
\subsubsection{Power-Scaling Law} 
\begin{figure}[t]
    \centerline{\includegraphics[width=0.8\textwidth]{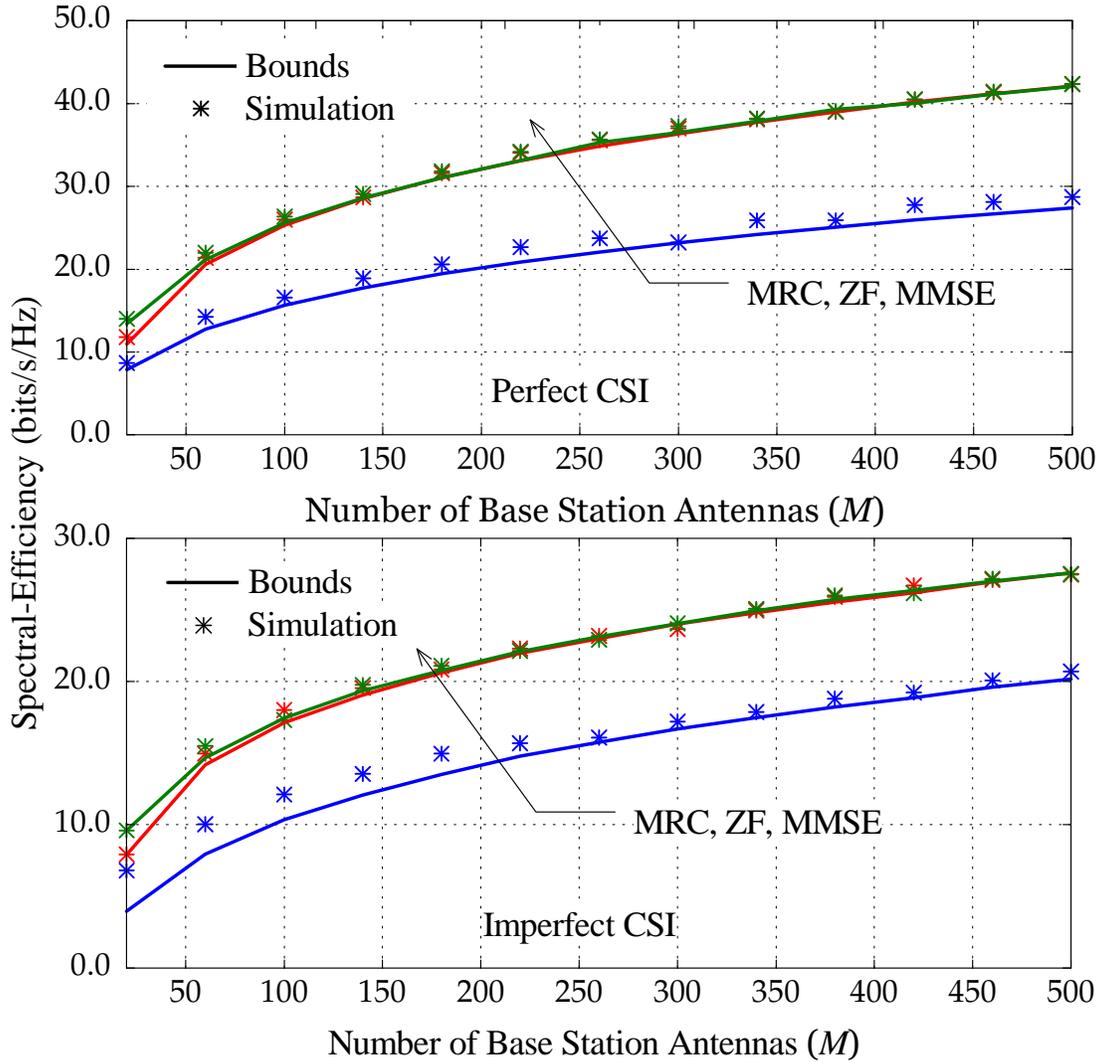}}
    \caption{Lower bounds and numerically evaluated values of the spectral efficiency for
    different  numbers of
BS antennas for MRC, ZF, and MMSE with perfect and imperfect CSI.
In this example there are  $K=10$ users, the coherence interval
$T=196$, the transmit power per terminal is $\Pu=10$ dB, and the
propagation channel parameters were $\sigma_{\mathrm{shadow}} = 8$
dB, and $\nu = 3.8$.}
    \label{fig:1}
\end{figure}

\label{sec:
Numerical Power} We first conduct an experiment to   validate the
tightness of our proposed capacity bounds. Fig.~\ref{fig:1} shows
the simulated spectral efficiency and the proposed analytical
bounds for MRC, ZF, and MMSE receivers with perfect and imperfect
CSI at $\Pu = 10$ dB. In this example there are $K=10$ users. For
CSI estimation from uplink pilots, we choose pilot sequences of
length $\tau=K$. (This is the smallest amount of training that can
be used.)
 Clearly, all bounds are very
tight, especially at large $M$. Therefore, in the following, we
will use these bounds for all numerical work.

\begin{figure}[t]
    \centerline{\includegraphics[width=0.8\textwidth]{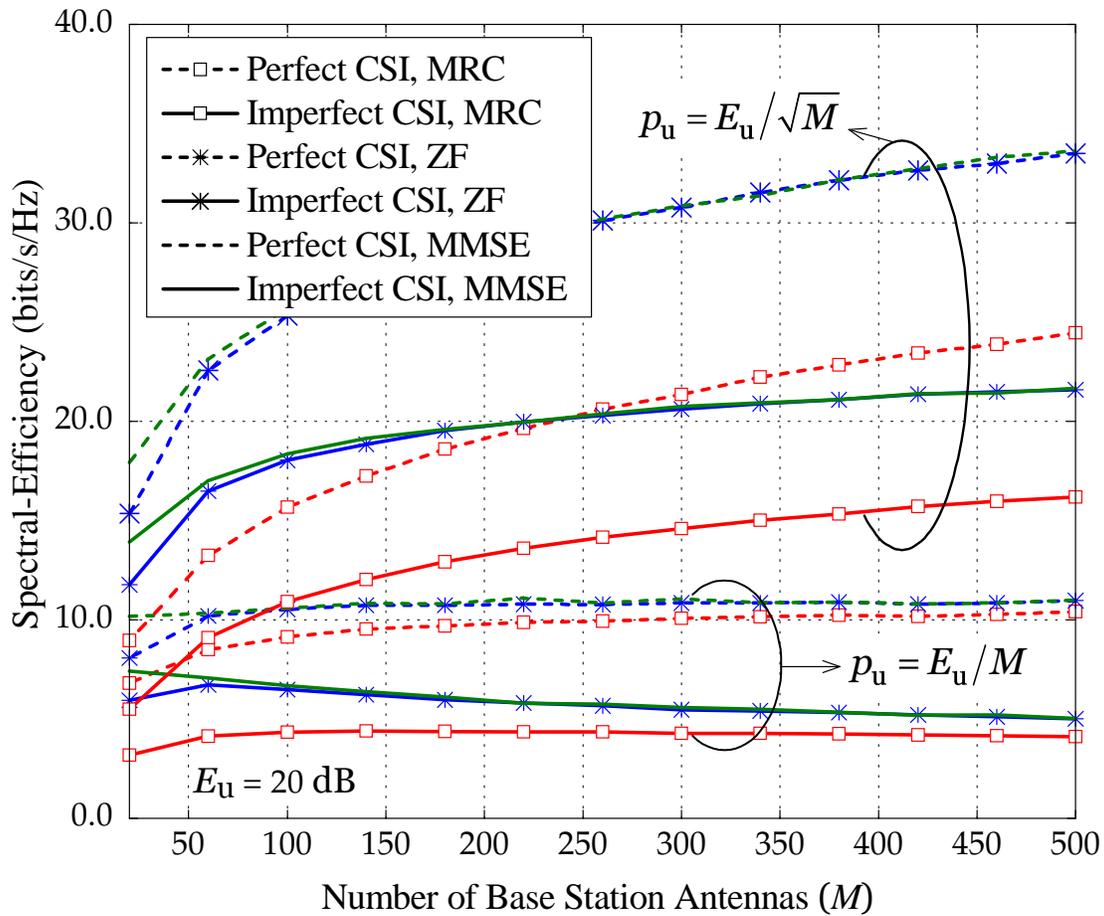}}
    \caption{Spectral efficiency versus the number of
BS antennas $M$ for MRC, ZF, and MMSE processing at the receiver,
with perfect CSI and with imperfect CSI (obtained from uplink
pilots). In this example  $K=10$ users are served simultaneously,
the reference transmit power is $\Eu=20$ dB, and the propagation
parameters were $\sigma_{\mathrm{shadow}} = 8$ dB and $\nu =
3.8$.}
    \label{fig:2}
\end{figure}

We next illustrate the power scaling laws. Fig.~\ref{fig:2} shows
the spectral efficiency on the uplink versus the number of BS
antennas for $\Pu = \Eu/M$ and $\Pu = \Eu/\sqrt{M}$ with perfect
and imperfect receiver CSI, and with MRC, ZF, and MMSE processing,
respectively. Here, we choose $\Eu = 20$ dB. At this SNR, the
spectral efficiency is in the order of 10--30 bits/s/Hz,
corresponding to a spectral efficiency per user of 1--3 bits/s/Hz.
These operating points are reasonable from a practical point of
view. For example, 64-QAM with a rate-1/2 channel code would
correspond to 3 bits/s/Hz. (Figure~\ref{fig:3}, see below, shows
results at lower SNR.) As expected, with $\Pu=\Eu/M$, when $M$
increases, the spectral efficiency approaches a constant value for
the case of perfect CSI, but decreases to $0$ for the case of
imperfect CSI. However, with $\Pu=\Eu/\sqrt{M}$, for the case of
perfect CSI the spectral efficiency grows without bound
(logarithmically fast with $M$) when $M\to\infty$ and with
imperfect CSI, the spectral efficiency converges to a nonzero
limit as $M \to\infty$.  These results confirm that we can scale
down the transmitted power of each user as $\Eu/M$ for the perfect
CSI case, and as $\Eu/\sqrt{M}$ for the imperfect CSI case when
$M$ is large.

Typically ZF is better than MRC at high SNR, and vice versa at low
SNR \cite{TV:05:Book}.   MMSE always performs the best across the
entire SNR range (see Remark~\ref{re: MMSE}). When comparing MRC
and ZF in Fig.~\ref{fig:2}, we see that here, when the transmitted
power is proportional to $1/\sqrt{M}$, the power is not low enough
to make MRC perform as well as ZF. But when the transmitted power
is   proportional to $1/M$,   MRC performs almost as well as ZF
for large $M$. Furthermore, as we can see from the figure,  MMSE
is always better than MRC or ZF, and its performance is very close
to ZF.

In Fig.~\ref{fig:3}, we consider the same setting as in
Fig.~\ref{fig:2}, but we choose $\Eu =5$ dB. This figure provides
the same insights as Fig.~\ref{fig:2}.  The gap between the
performance of MRC and that of  ZF (or MMSE) is reduced compared
with Fig.~\ref{fig:2}. This is so because the relative effect of
crosstalk interference (the interference from other users) as
compared to the thermal noise is smaller here than in
Fig.~\ref{fig:2}.

\begin{figure}[t]
\centerline{\includegraphics[width=0.8\textwidth]{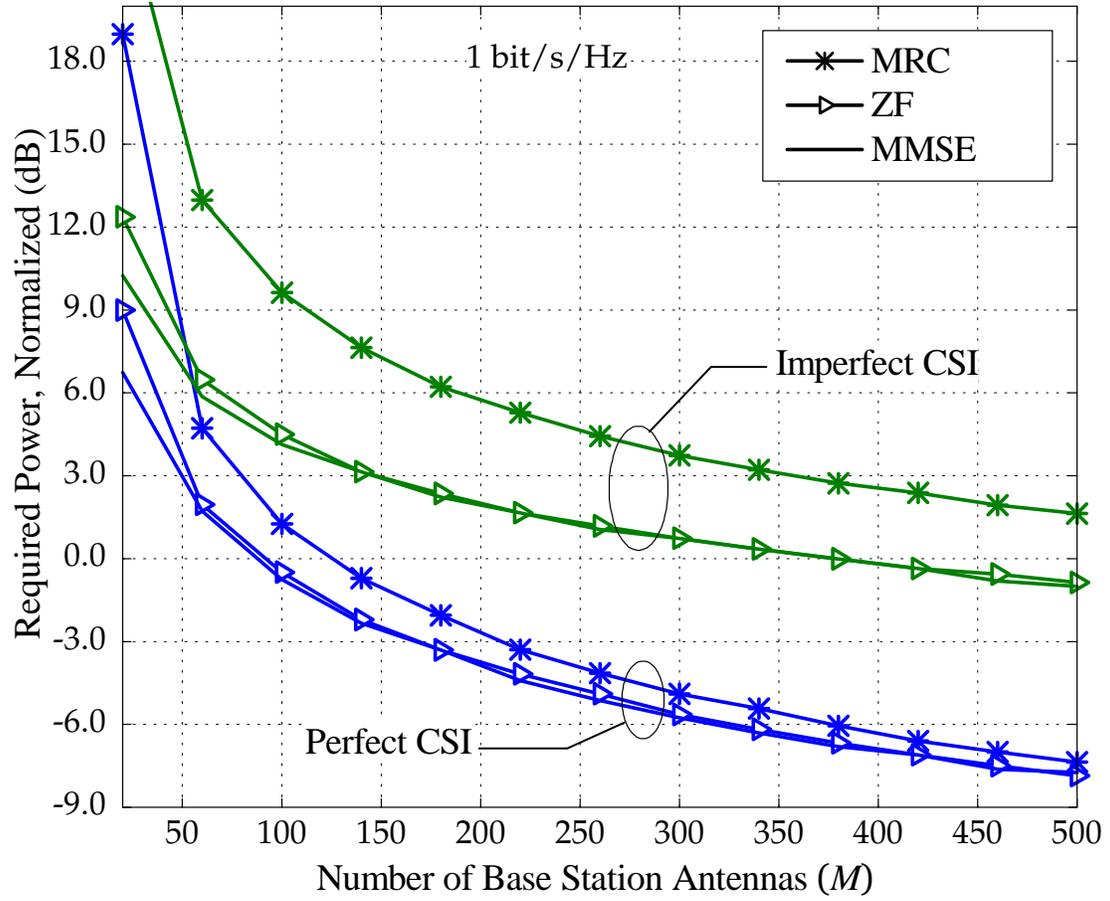}}
\caption{Transmit power required to achieve  $1$ bit/channel use
per user for MRC, ZF, and MMSE processing, with perfect and
imperfect CSI, as a function of the number $M$ of BS antennas. The
number of users is fixed to  $K= 10$, and the propagation
parameters are $\sigma_{\mathrm{shadow}} = 8$ dB  and $\nu =
3.8$.} \label{fig:4}
\end{figure}

\begin{figure}[t]
\centerline{\includegraphics[width=0.8\textwidth]{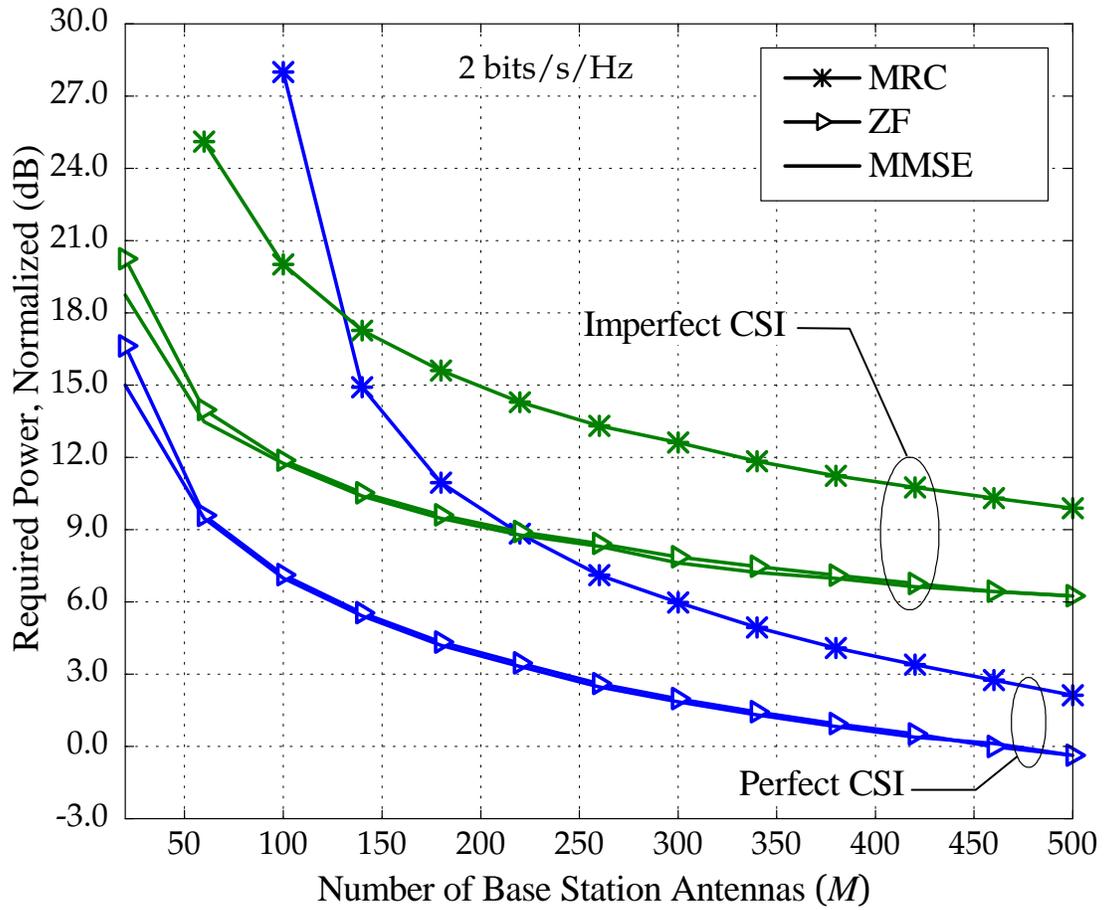}}
\caption{Same as Figure~\ref{fig:4} but for a target spectral
efficiency of 2 bits/channel use per user. } \label{fig:5}
\end{figure}

We next show the  transmit power per  user that is needed to reach
a fixed spectral efficiency. Fig.~\ref{fig:4} shows the normalized
power ($\Pu$) required to achieve $1$ bit/s/Hz per user as a
function of  $M$.  As predicted by the analysis, by doubling $M$,
we can cut back the power by approximately 3 dB and 1.5 dB for the
cases of  perfect   and imperfect CSI, respectively. When $M$ is
large ($M/K \gtrapprox 6$), the difference in performance between
MRC and ZF (or MMSE) is less than $1$ dB and $3$ dB for the cases
of perfect and imperfect CSI, respectively. This difference
increases when we increase the target spectral efficiency.
Fig.~\ref{fig:5}  shows the normalized power required for $2$
bit/s/Hz per user. Here, the crosstalk interference is more
significant (relative to the thermal noise) and hence the ZF and
MMSE receivers  perform relatively better.

%\vspace{-0.5cm}
\subsubsection{Energy Efficiency versus Spectral
Efficiency Tradeoff }

We next examine the tradeoff between energy efficiency and
spectral efficiency in more detail. Here, we ignore the effect of
large-scale fading, i.e., we set $\B{D}=\B{I}_K$. We normalize the
energy efficiency against a reference mode corresponding to a
single-antenna BS serving one single-antenna user with $\Pu = 10$
dB. For this reference mode, the spectral efficiencies and energy
efficiencies for MRC, ZF, and MMSE are equal, and given by (from
\eqref{MRC IPCSI 2} and \eqref{eq: ES 1})
%$$R_{\mathrm{P}}^{0} = \E\left\{\log_2\left(1 + \Pu |z|^2\right) \right\}, ~ \eta_{\mathrm{P}}^{0} = R_{\mathrm{P}}^{0}/\Pu, ~ \text{for perfect SCI}$$
$$R_{\mathrm{IP}}^{0} = \frac{T-\tau}{T}\E\left\{\log_2\left(1 + \frac{\tau\Pu^2 |z|^2}{1 + \Pu\left(1+\tau\right)}\right) \right\},
~ \eta_{\mathrm{IP}}^{0} = R_{\mathrm{IP}}^{0}/\Pu$$ where $z$ is
a Gaussian RV with zero mean and unit variance. For the reference
mode, the spectral-efficiency is obtained by choosing the duration
of the uplink pilot sequence $\tau$ to maximize
$R_{\mathrm{IP}}^{0}$. Numerically we find that
$R_{\mathrm{IP}}^{0} = 2.65$ bits/s/Hz and $\eta_{\mathrm{IP}}^{0}
= 0.265$ bits/J.

\begin{figure}[t]
\centerline{\includegraphics[width=0.8\textwidth]{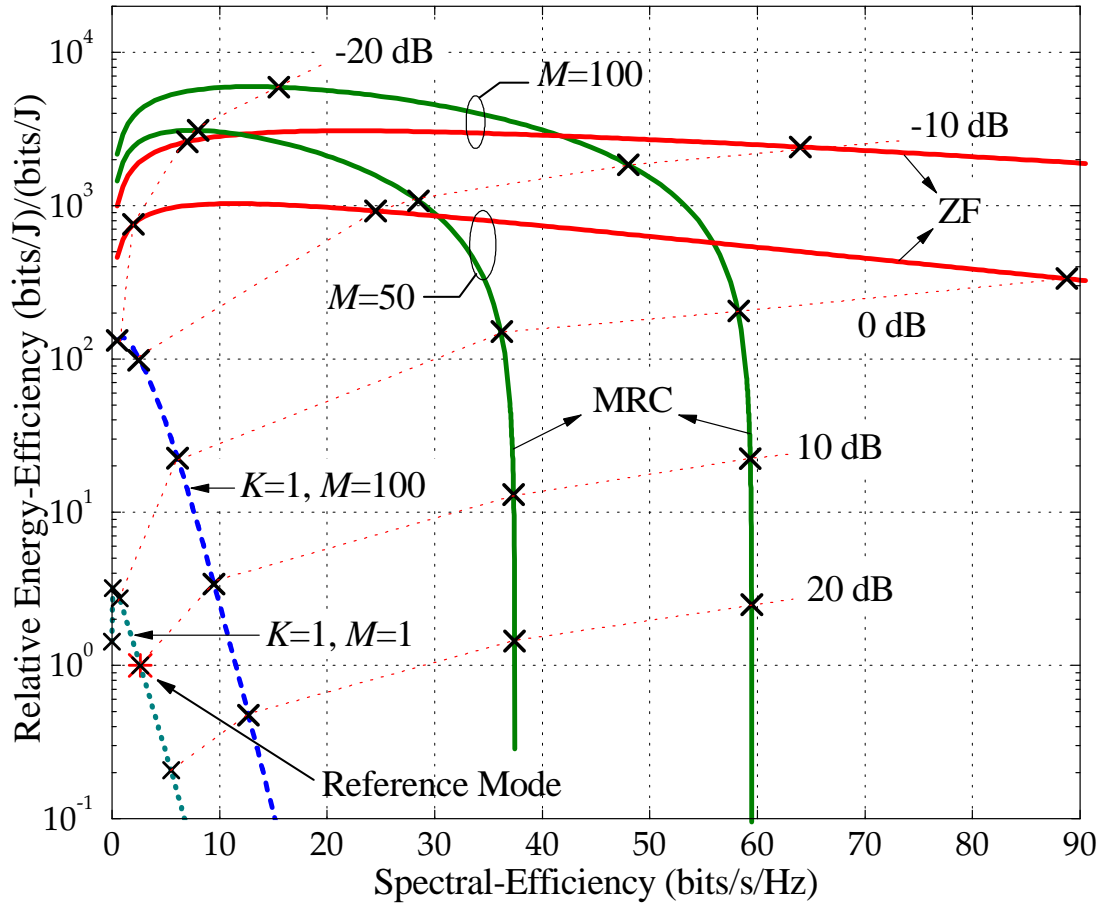}}
\caption{Energy efficiency (normalized with respect to the
reference mode) versus spectral efficiency for MRC and ZF receiver
processing with imperfect CSI. The reference mode corresponds to
$K=1, M=1$ (single antenna, single user), and a transmit power of
$\Pu=10$ dB. The coherence interval is $T=196$ symbols. For the
dashed curves (marked with $K=1$), the transmit power $\Pu$ and
the fraction of the coherence interval $\tau/T$ spent on training
was optimized in order to maximize the energy efficiency for a
fixed spectral efficiency.  For the green and red curves (marked
MRC and ZF; shown for $M=50$ and $M=100$ antennas, respectively),
the number of users $K$ was optimized jointly with $\Pu$ and
$\tau/T$ to maximize the energy efficiency for given spectral
efficiency. Any operating point on the curves can be obtained by
appropriately selecting $\Pu$ and optimizing with respect to $K$
and $\tau/T$.  The number marked next to the $\times$ marks on
each curve is the power $\Pu$ spent by the transmitter. }
\label{fig:6}
\end{figure}

Fig.~\ref{fig:6} shows the relative energy efficiency versus the
the spectral efficiency for MRC and ZF. The relative energy
efficiency is obtained by normalizing the energy efficiency by
$\eta_{\mathrm{IP}}^{0}$ and it is therefore dimensionless. The
dotted  and dashed lines show the performances for the cases of
$M=1, K=1$ and $M=100, K=1$, respectively. Each point on the
curves is obtained by choosing the transmit power $\Pu$ and pilot
sequence length $\tau$ to maximize the energy efficiency for a
given spectral efficiency. The solid lines show the performance
for the cases of $M=50$, and $100$. Each point on these curves is
computed by jointly choosing $K$, $\tau$, and $\Pu$ to maximize
the energy-efficiency subject a fixed  spectral-efficiency, i.e.,
\begin{align*}
%    %\left(%
%\begin{array}{l}
%  \arg \mathop {\max}\limits_{\Pu, K, \tau} ~ \eta_{\mathrm{IP}}^{\tt A} \\
%  \hspace{1.7cm} \text{s.t.} ~ R_{\mathrm{IP}}^{\tt A} = \text{const.}, K \leq \tau \leq T \\
%\end{array}%
%%\right)
\arg \mathop {\max}\limits_{\Pu, K, \tau} ~
\eta_{\mathrm{IP}}^{\tt A}, \hspace{1.0cm} \text{s.t.} ~
R_{\mathrm{IP}}^{\tt A} = \text{const.}, K \leq \tau \leq T
\end{align*}

We first consider a single-user system with $K=1$.  We compare the
performance of the cases $M=1 $  and $M=100 $. Since $K=1$  the
performances of MRC and ZF are equal. With the same power used as
in the reference mode, i.e., $\Pu=10$ dB, using $100$ antennas can
increase the spectral efficiency and the energy efficiency by
factors of $4$ and $3$, respectively. Reducing the transmit power
by a factor of $100$, from $10$ dB to $-10$ dB yields a $100$-fold
improvement in energy efficiency compared with that of the
reference mode with no reduction in spectral-efficiency.

We next consider a multiuser system ($K>1$).  Here
 the transmit power $\Pu$, the number of
users $K$, and the duration of pilot sequences $\tau$ are chosen
optimally for fixed $M$. We consider $M=50$ and $100$. Here  the
system performance improves very significantly compared to the
single-user case. For example, with MRC, at $\Pu=0$ dB, compared
with the case of $M=1, K=1$, the spectral-efficiency increases by
factors of $50$ and $80$, while the energy-efficiency increases by
factors of $55$ and $75$ for $M=50$ and $M=100$, respectively. As
discussed in Section~\ref{sec: Energy-Spectral}, at low spectral
efficiency, the energy efficiency increases when the spectral
efficiency increases. Furthermore, we can see that at high
spectral efficiency, ZF outperforms MRC. This is due to the fact
that the MRC receiver is limited by the intracell interference,
which is significant   at high spectral efficiency. As a
consequence, when $\Pu$ is increased, the spectral efficiency of
MRC approaches  a constant value, while the energy efficiency goes
to zero (see \eqref{eq: ES MRC3}).%\footnote{ Note that, certain
%activities (multiplexing to many users rather that beam-forming to
%a single user and increasing the number of service antennas) do
%simultaneously benefit both the spectral-efficiency and the
%radiated energy-efficiency. The point is that, once the number of
%service-antennas is set, one can adjust other system parameters
%(radiated power, numbers of users, duration of pilot sequences) to
%obtain increased spectral-efficiency at the cost of reduced
%energy-efficiency, and vice-versa. In fact this should be a
%desirable feature for service providers: they can set the
%operating point according to the current traffic demand (high
%energy-efficiency and low spectral-efficiency, for example, during
%periods of low demand).}

\begin{figure}[t]
\centerline{\includegraphics[width=0.8\textwidth]{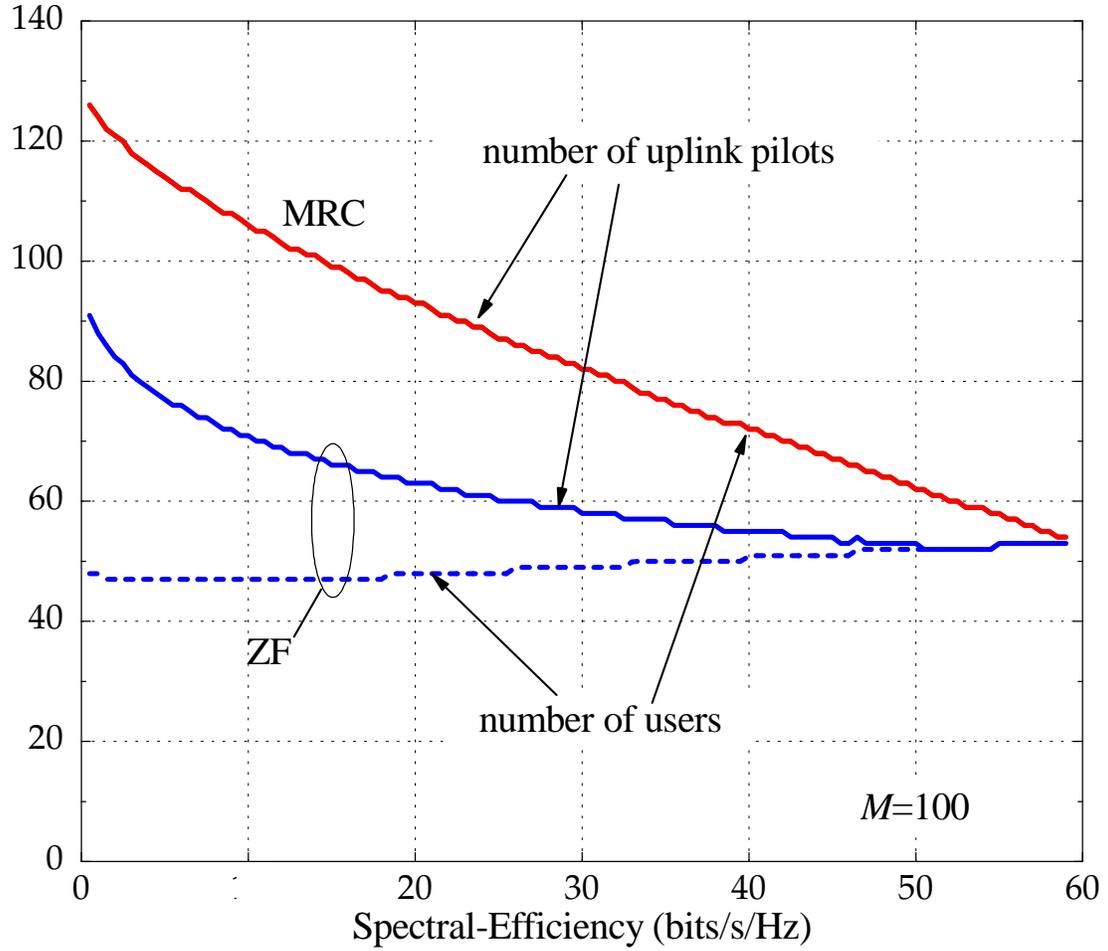}}
\caption{Optimal number of users $K$ and number of symbols $\tau$
spent on training, out of a total of $T=196$ symbols per coherence
interval, for the curves in  Fig.~\ref{fig:6} corresponding to
$M=100$ antennas. } \label{fig:7}
\end{figure}

The corresponding optimum values of $K$ and  $\tau$ as functions
of the spectral efficiency for $M=100$ are shown in
Fig.~\ref{fig:7}. For MRC, the optimal number of users and uplink
pilots are the same (this means that the minimal possible length
of  training sequences are used). For ZF, more of the coherence
interval is used for training. Generally, at low transmit power
and therefore at low spectral efficiency, we spend more time on
training than on payload data transmission.  At high power (high
spectral efficiency and low energy efficiency), we can serve
around $55$ users, and $K=\tau$ for both MRC and ZF.

\vspace{-0.5cm}
\subsection{Multicell MU-MIMO Systems}

Next, we examine the effect of pilot contamination on the energy
and spectral efficiency for multicell systems. We consider a
system with $L=7$ cells. Each cell has the same size as in the
single-cell system. When shrinking the cell size, one typically
also cuts back on the power. Hence, the relation between signal
and interference power would not be substantially different in
systems with smaller cells and in that sense, the analysis is
largely independent of the actual physical size of the cell
\cite{LHA:12:Arxiv}. Note that, setting $L=7$ means that we
consider the performance of a given cell with the interference
from $6$ nearest-neighbor cells. We assume $\B{D}_{ll}=\B{I}_K$,
and $\B{D}_{li}=\beta \B{I}_K$, for $i\neq l$. To examine the
performance in a practical scenario, the intercell interference
factor, $\beta$, is chosen as follows. We consider two users, the
$1$st user is located  uniformly at random in the first cell, and
the $2$nd user is located uniformly at random in one of the $6$
nearest-neighbor cells of the $1$st cell. Let $\bar{\beta}_1$ and
$\bar{\beta}_2$ be the large scale fading from the $1$st user and
the $2$nd user to the $1$st BS, respectively. (The large scale
fading is modelled as in Section~\ref{sec: Numerical Power}.) Then
we compute $\beta$ as
$\E\left\{\bar{\beta}_2/\bar{\beta}_1\right\}$. By simulation, we
obtain $\beta=0.32, 0.11$, and $0.04$ for the cases of
($\sigma_{\mathrm{shadow}}=8$ dB, $\nu = 3.8, f_{\mathrm{reuse}} =
1$), ($\sigma_{\mathrm{shadow}}=8$ dB, $\nu = 3,
f_{\mathrm{reuse}} = 1$), and ($\sigma_{\mathrm{shadow}}=8$ dB,
$\nu = 3.8, f_{\mathrm{reuse}} = 3$), respectively, where
$f_{\mathrm{reuse}}$ is the frequency reuse factor.

\begin{figure}[t]
\centerline{\includegraphics[width=0.8\textwidth]{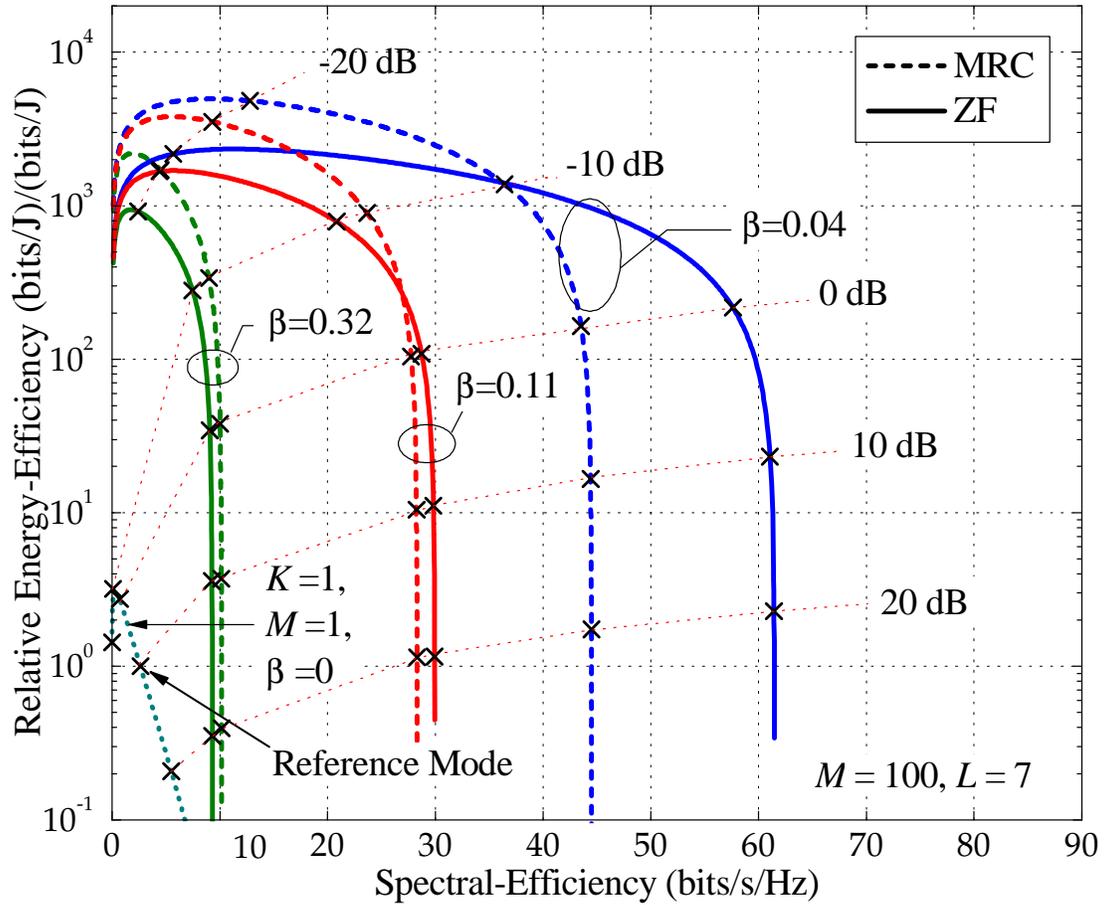}}
\caption{Same as Figure~\ref{fig:6}, but for a multicell scenario,
with $L=7$ cells, and coherence interval $T = 196$.} \label{fig:8}
\end{figure}

Fig.~\ref{fig:8} shows the relative energy efficiency versus the
spectral efficiency for MRC and ZF of the multicell system. The
reference mode is the same as the one in Fig.~\ref{fig:6} for a
single-cell system. The dotted line shows the performance for the
case of $M = 1, K=1$, and $\beta =0$. The solid and dashed lines
show the performance for the cases of $M=100$, and $L=7$, with
different intercell interference factors $\beta$ of $0.32, 0.11$,
and $0.04$. Each point on these curves is computed by jointly
choosing $\tau$, $K$, and $\Pu$ to maximize the energy efficiency
for a given spectral efficiency. We can see that the pilot
contamination significantly degrades the system performance. For
example, when $\beta$ increases from $0.11$ to $0.32$ (and hence,
the pilot contamination increases), with the same power, $\Pu =
10$ dB, the spectral efficiency and the energy efficiency reduce
by factors of $3$ and $2.7$, respectively. However, with low
transmit power where the spectral efficiency is smaller than $10$
bits/s/Hz, the system performance is not affected much by the
pilot contamination. Furthermore, we can see that in a multicell
scenario with high pilot contamination, MRC achieves a better
performance than ZF.

\section{Conclusion} \label{Sec:Conclusion}

Very large MIMO systems offer the opportunity of increasing the
spectral efficiency (in terms of bits/s/Hz sum-rate in a given
cell) by one or two orders of magnitude, and simultaneously
improving the energy efficiency (in terms of bits/J) by three
orders of magnitude. This is possible with simple linear
processing such as MRC or ZF at the BS, and using channel
estimates obtained from uplink pilots even in a high mobility
environment where half of the channel coherence interval is used
for training. Generally, ZF outperforms MRC owing to its ability
to cancel intracell interference. However, in multicell
environments with strong pilot contamination, this advantage tends
to diminish. MRC has the additional benefit of facilitating a
distributed per-antenna implementation of the detector. These
conclusions are valid in an operating regime where $100$ antennas
serve about $50$ terminals in the same time-frequency resource,
each terminal having a fading-free throughput of about $1$ bpcu,
and hence the system offering a sum-throughput of about $50$ bpcu.

%We studied the uplink transmission of data from $K$ terminals to
%an  array of $M$ antennas at the BS, with respect to energy
%efficiency and throughput. We found that energy efficiency is
%qualitatively different depending on whether the receiver has
%perfect  CSI  or whether it has only imperfect CSI  derived from
%uplink pilots. With perfect CSI the radiated power of the
%terminals can be made inversely proportional to $M$ while
%maintaining spatial multiplexing gains, while for imperfect CSI
%the power can only be made inversely proportional to the
%square-root of $M$. We also compared three receiver structures:
%MRC, ZF, and MMSE. Among these detectors, the MMSE is always the
%best one, and its performance is very close to that of ZF. Except
%for the case of imperfect channel knowledge and a small number of
%antennas, MMSE and ZF outperform MRC. However,  reducing the
%transmit power  narrows the performance gap between MRC and ZF (or
%MMSE).
%
%Furthermore, we investigated the tradeoff between spectral
%efficiency and energy efficiency. We showed that, at low spectral
%efficiency, improving spectral efficiency also can improve the
%energy efficiency. Taken together, a large excess of antennas
%offers high spectral efficiency, high energy efficiency, with
%low-complexity processing.

% use section* for acknowledgement
\appendix

\subsection{Proof of Proposition~\ref{Prop PCSI MCR}} \label{app:1}
From \eqref{eq: LBRate PCSI 1a}, we have
\begin{align} \label{eq: GRate PCSI 5b}
    \tilde{R}_{\mathrm{P},k}^{\tt{mrc}}
    =
       \log_2
            \left(
            1 +
        \left(
            \E
            \left\{
            \frac{
                \Gu \sum_{i=1, i \neq k}^{K}
                | \tilde{g}_i|^2
                + 1
                }{
            \Gu \| \B{g}_k \|^2
                }
            \right\}
            \right)^{-1}
            \right)
\end{align}
where $\tilde{g}_i \triangleq \frac{\B{g}_k^H
\B{g}_i}{\|\B{g}_k\|}$. Conditioned on $\B{g}_k$, $\tilde{g}_i$ is
a Gaussian RV with zero mean and variance $\beta_i$ which does not
depend on $\B{g}_k$. Therefore, $\tilde{g}_i$ is Gaussian
distributed and independent of $\B{g}_k$, $\tilde{g}_i \sim
\CG{0}{\beta_i}$. Then,
\begin{align}\label{eq app1 1}
            \E
            \left\{
            \frac{
                \Gu \sum_{i=1, i \neq k}^{K}
                | \tilde{g}_i|^2
                + 1
                }{
            \Gu \| \B{g}_k \|^2
                }
            \right\}
            &=
                \left(
                \Gu \sum_{i=1, i \neq k}^{K}
                \E \left\{| \tilde{g}_i|^2\right\}
                + 1
                \right)
                \E \left\{
            \frac{
                1
                }{
                \Gu\| \B{g}_k \|^2
                }
                \right\}
            \nonumber
            \\
            &=
                \left(
                \Gu \sum_{i=1, i \neq k}^{K}
                \beta_i
                + 1
                \right)
                \E \left\{
            \frac{
                1
                }{
                \Gu\| \B{g}_k \|^2
                }
                \right\}
\end{align}
Using the identity \cite{TV:04:FTCIT}
\begin{align}\label{eq app1 2}
    \E \left\{
        {\tt tr} \left(\B{W}^{-1} \right)
        \right\}
    =
    \frac{m}{n-m}
\end{align}
where $\B{W} \sim \mathcal{W}_{m}\left(n, \B{I}_n \right)$ is an
$m\times m$ central complex Wishart matrix with $n$ ($n > m$)
degrees of freedom, we obtain
\begin{align}\label{eq app1 4}
                \E \left\{
            \frac{
                1
                }{
                \Gu\| \B{g}_k \|^2
                }
                \right\}
        &=
            \frac{1}
                {\Gu \left(M-1 \right)\beta_k}, ~ \text{for $M \geq 2$}
\end{align}
Substituting \eqref{eq app1 4} into \eqref{eq app1 1}, we arrive
at the desired result \eqref{eq: LBRate PCSI 1}.

\subsection{Proof of Proposition~\ref{Prop PCSI ZF}} \label{app:2}
From \eqref{eq CM 2}, we have
\begin{align} \label{Proof lemma2 3}
    \E  \left\{
        \left[\left(\B{G}^H \B{G}\right)^{-1}\right]_{kk}
        \right\}
    &=
        \frac{1}{\beta_k}
        \E  \left\{
        \left[\left(\B{H}^H \B{H}\right)^{-1}\right]_{kk}
        \right\}
    =
        \frac{1}{K \beta_k}
        \E  \left\{{\tt tr}
        \left[\left(\B{H}^H \B{H}\right)^{-1}\right]
        \right\}
%    \nonumber
%    \\
    \nonumber
    \\
    &\mathop  = \limits^{(a)}
        \frac{1}{\left(M-K\right)\beta_k}, ~ \text{for $M \geq K+1$}
\end{align}
where $(a)$ is obtained by using \eqref{eq app1 2}. Using
\eqref{Proof lemma2 3}, we get \eqref{ZFRate PCSI 2b}.

 %\bibliographystyle{IEEEtran}
% \bibliography{IEEEabrv,CCTLABBiblio}

\begin{thebibliography}{10}
\providecommand{\url}[1]{#1} \csname url@samestyle\endcsname
\providecommand{\newblock}{\relax}
\providecommand{\bibinfo}[2]{#2}
\providecommand{\BIBentrySTDinterwordspacing}{\spaceskip=0pt\relax}
\providecommand{\BIBentryALTinterwordstretchfactor}{4}
\providecommand{\BIBentryALTinterwordspacing}{\spaceskip=\fontdimen2\font
plus \BIBentryALTinterwordstretchfactor\fontdimen3\font minus
  \fontdimen4\font\relax}
\providecommand{\BIBforeignlanguage}[2]{{%
\expandafter\ifx\csname l@#1\endcsname\relax
\typeout{** WARNING: IEEEtran.bst: No hyphenation pattern has been}%
\typeout{** loaded for the language `#1'. Using the pattern for}%
\typeout{** the default language instead.}%
\else \language=\csname l@#1\endcsname \fi #2}}
\providecommand{\BIBdecl}{\relax} \BIBdecl

\vspace{-0.2cm}
\bibitem{NLM:11:ACCCC}
H.~Q. Ngo, E.~G. Larsson, and T.~L. Marzetta, ``Uplink power
efficiency of
  multiuser {MIMO} with very large antenna arrays,'' in \emph{Proc.
  Allerton Conf. Commun., Control, Comput.},
  Urbana-Champaign, IL., Sept. 2011, pp.1272-1279.

\bibitem{GKHCS:07:SPM}
D.~Gesbert, M.~Kountouris, R.~W. {Heath Jr.}, C.-B. Chae, and
T.~S{\"{a}}lzer,
  ``Shifting the {MIMO} paradigm,'' \emph{IEEE Sig.\ Proc.\ Mag.}, vol.~24,
  no.~5, pp. 36--46, 2007.

\bibitem{CJKR:10:IT}
G.~Caire, N.~Jindal, M.~Kobayashi, and N.~Ravindran, ``Multiuser
{MIMO}
  achievable rates with downlink training and channel state feedback,''
  \emph{{IEEE} Trans. Inf. Theory}, vol.~56, no.~6, pp. 2845--2866, 2010.

%\bibitem{JAMV:09:ISIT}
%J.~Jose, A.~Ashikhmin, T.~L. Marzetta, and S.~Vishwanath, ``Pilot
%contamination
%  problem in multi-cell {TDD} systems,'' in \emph{Proc. IEEE International
%  Symposium on Information Theory (ISIT'09)}, Seoul, Korea, Jun. 2009, pp.
%  2184--2188.

\bibitem{JAMV:11:WCOM}
J.~Jose, A.~Ashikhmin, T.~L. Marzetta, and S.~Vishwanath, ``Pilot
contamination  and precoding in multi-cell {TDD} systems,''
\emph{{IEEE} Trans. Wireless Commun.}, vol.~10, no.~8, pp.
2640--2651, Aug. 2011.

\bibitem{Verdu:89:Al}
S.~Verd{\'u}, \emph{Multiuser Detection}, Cambridge University
Press, 1998.

\bibitem{VT:03:IT}
P.~Viswanath and D.~N.~C. Tse, ``Sum capacity of the vector
{G}aussian
  broadcast channel and uplink-downlink duality'' \emph{{IEEE} Trans. Inf.
  Theory}, vol.~49, no.~8, pp. 1912--1921, Aug. 2003.

\bibitem{WSS:06:IT}
H.~Weingarten, Y.~Steinberg, and S.~Shamai, ``The capacity region
of the
  Gaussian multiple-input multiple-output broadcast channel,'' \emph{{IEEE}
  Trans. Inf. Theory}, vol.~52, no.~9, pp. 3936--3964, Sep. 2006.

\bibitem{Mar:10:WCOM}
T.~L. Marzetta, ``Noncooperative cellular wireless with unlimited
numbers of
  BS antennas,'' \emph{{IEEE} Trans. Wireless Commun.}, vol.~9,
  no.~11, pp. 3590--3600, Nov. 2010.

\bibitem{Mar:06:ACSSC}
------, ``How much training is required for multiuser {MIMO},'' in
  \emph{Fortieth Asilomar Conference on Signals, Systems and Computers (ACSSC
  '06)}, Pacific Grove, CA, USA, Oct. 2006, pp. 359--363.

\bibitem{RPLLMET:11:SPM}
\BIBentryALTinterwordspacing F.~Rusek, D.~Persson, B.~K. Lau,
E.~G. Larsson, T.~L. Marzetta, O.~Edfors, and
  F.~Tufvesson, ``Scaling up {MIMO}: Opportunities and challenges with very
  large arrays,'' \emph{IEEE Sig.\ Proc.\ Mag.}, accepted. [Online]. Available:
  {arxiv.org/abs/1201.3210}.
\BIBentrySTDinterwordspacing

\bibitem{HBD:11:ACCCC}
J. Hoydis, S.~ten Brink, and M. Debbah, ``Massive {MIMO}: How many
antennas do we need?,'' in \emph{Proc. 49th
  Allerton Conference on Communication, Control, and Computing}, 2011.


 \bibitem{ComMag}
 A.\ Fehske, G.\ Fettweis, J.\ Malmodin and G.\ Biczok,
 ``The global footprint of mobile communications: the ecological and economic perspective,''
 {\em IEEE Communications Magazine}, pp.\ 55-62, August 2011.

\bibitem{TV:05:Book}
D.~N.~C. Tse and P.~Viswanath, \emph{Fundamentals of Wireless
  Communications}.\hskip 1em plus 0.5em minus 0.4em\relax Cambridge, UK:
  Cambridge University Press, 2005.

\bibitem{HCPR:11:APWC}
H. Huh, G. Caire, H. C. Papadopoulos, S. A. Rampshad, ``Achieving
large spectral efficiency with {TDD} and not-so-many base station
antennas,'' in \emph{Proc. IEEE Antennas and Propagation in
Wireless Communications (APWC)}, 2011.

\bibitem{WCSD:10:SPAWC}
S. Wagner, R. Couillet, D. T. M. Slock, and M. Debbah, ``Large
system analysis of zero-forcing precoding in MISO broadcast
channels with limited feedback,'' in \emph{Proc. IEEE Int. Works.
Signal Process. Adv. Wireless Commun. (SPAWC)}, 2010.

\bibitem{YM:12:JSAC}
H. Yang and T. L. Marzetta, ``Performance of conjugate and
zero-forcing beanforming in large-scale antenna systems'',
\emph{IEEE J. Select. Areas Commun.}, 2012, submitted.

\bibitem{Cra:70:Book}
H.~Cram\'{e}r, \emph{Random Variables and Probability
Distributions}.\hskip 1em
  plus 0.5em minus 0.4em\relax Cambridge, UK: Cambridge University Press, 1970.

%\bibitem{HH:03:IT}
%B.~Hassibi and B.~M. Hochwald, ``How much training is needed in
%  multiple-antenna wireless links?'' \emph{{IEEE} Trans. Inf. Theory}, vol.~49,
%  no.~4, pp. 951--963, Apr. 2003.

\bibitem{KP:08:WCOM}
N.~Kim and H.~Park, ``Performance analysis of {MIMO} system with
linear {MMSE}
  receiver,'' \emph{{IEEE} Trans. Wireless Commun.}, vol.~7, no.~11, pp.
  4474--4478, Nov. 2008.

\bibitem{GSC:98:COM}
H.~Gao, P.~J. Smith, and M.~Clark, ``Theoretical reliability of
MMSE linear
  diversity combining in Rayleigh-fading additive interference channels,''
  \emph{{IEEE} Trans. Commun.}, vol.~46, no.~5, pp. 666--672, May 1998.

\bibitem{LPNC:06:IT}
P.~Li, D.~Paul, R.~Narasimhan, and J.~Cioffi, ``On the
distribution of {SINR}
  for the {MMSE} {MIMO} receiver and performance analysis,'' \emph{{IEEE}
  Trans. Inf. Theory}, vol.~52, no.~1, pp. 271--286, Jan. 2006.

\bibitem{GR:07:Book}
I.~S. Gradshteyn and I.~M. Ryzhik, \emph{Table of Integrals,
Series, and
  Products}, 7th~ed.\hskip 1em plus 0.5em minus 0.4em\relax San Diego, CA:
  Academic, 2007.

\bibitem{TV:04:FTCIT}
A.~M. Tulino and S.~Verd\'{u}, ``Random matrix theory and wireless
  communications,'' \emph{Foundations and Trends in Communications and
  Information Theory}, vol.~1, no.~1, pp. 1--182, Jun. 2004.

\bibitem{LHA:12:Arxiv}
A.~Lozano, R.~W. Heath Jr., and J.~G. Andrews, ``Fundamentral
limits of cooperation,'' Mar. 2012. [Online]. Available:
  \url{arxiv.org/abs/1204.0011}.

%\bibitem{GHP:02:CL}
%D.~A. Gore, R.~W. {Heath Jr.}, and A.~J. Paulraj, ``Transmit
%antenna selection in
%  spatial multiplexing systems,'' \emph{{IEEE} Commun. Lett.}, vol.~6, no.~11,
%  pp. 491--493, Nov. 2002.

\end{thebibliography}

\clearpage

\begin{table}[h]
    \caption{
        Lower bounds on the achievable rates of the uplink transmission for the $k$th user.
    }
    \centerline{
\begin{tabular}{|c|c|c|}
  \hline
  % after \\: \hline or \cline{col1-col2} \cline{col3-col4} ...
   & Perfect CSI & Imperfect CSI \\
  \hline
  %\hline
  MRC & $ \log_2\left( 1 + \frac{ \Gu \left(M-1 \right) \beta_k
            }{
            \Gu
            \sum_{i=1, i \neq k}^{K}
            \beta_i
            +
            1
            }
    \right)$ &  $\log_2
            \left(
            1 +
            \frac{
                \tau \Gu^2
                \left(M-1 \right)
                \beta_k^2
                }{
                \Gu
                \left(\tau \Gu\beta_k +1 \right)
                \sum_{i=1, i \neq k}^{K} \beta_i
                +
                \left(\tau +1 \right)\Gu\beta_k
                +
                1
                }
            \right)$ \\ \hline
  ZF & $\log_2 \left(1 + \Gu \left(M-K\right) \beta_k \right)$ &
            $\log_2
            \left(
            1 +
            \frac{
                \tau \Gu^2 \left(M-K\right)
                \beta_k^2
                }{
                \left(
                \tau \Gu \beta_k
                +1
                \right)
                    \sum_{i=1}^{K}
                    \frac{\Gu\beta_i }{\tau \Gu \beta_i +1}
                +
                \tau \Gu \beta_k
                +1
                }
            \right)$ \\ \hline
  MMSE & $\log_2\left(1 + \left(\alpha_k -1 \right)\theta_k \right)$ & $\log_2\left(1 + \left(\hat{\alpha}_k -1 \right)\hat{\theta}_k \right)$ \\  \hline
\end{tabular}}
    \label{table:1}
\end{table}

\clearpage

\clearpage

\clearpage

\begin{figure}[t]
    \centerline{\includegraphics[width=0.8\textwidth]{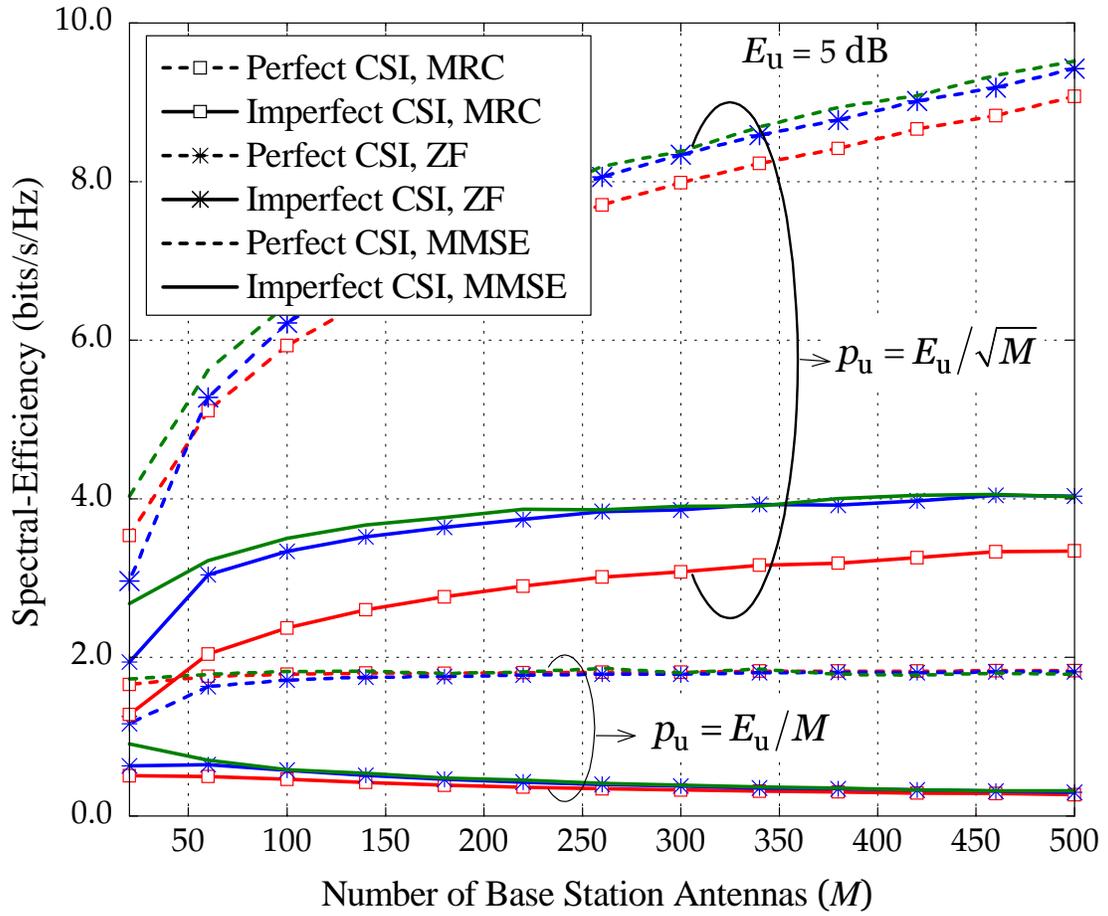}}
    \caption{Same as Figure~\ref{fig:2}, but with $\Eu = 5$ dB.}
    \label{fig:3}
\end{figure}

\clearpage

\clearpage

\clearpage

\clearpage

\clearpage

\end{document}